\documentclass[manuscript]{aastex}

\usepackage{graphicx}
\usepackage{epstopdf}
\usepackage{amsthm}
\usepackage{natbib}

\newcommand{\emdx}{{\bf EMDX}}
\newcommand{\viored}{{\bf VIORED}}
\newcommand{\kvk}{{\bf K2VK3}}
\newcommand{\delkmax}{{\bf DELK1}}
\newcommand{\delkmin}{{\bf DELK2}}
\newcommand{\delwb}{{\bf DELWB}}
\newcommand{\kthree}{{\bf K3}}

\begin{document}
\large 
\begin{center}

\vskip -1in
{\large\bf Solar Cycle Variability 
and Surface Differential Rotation from Ca II K-Line Time Series Data}
\vskip .2in
Jeffrey D. Scargle,
Space Science and Astrobiology Division\\
NASA Ames Research Center

Stephen L. Keil,
National Solar Observatory

Simon P. Worden,
NASA Ames Research Center

\end{center}


\noindent Draft of \today.
Note: this version makes use of slightly more
data than that submitted to the Astrophysical 
Journal, extending to February 28, 2013.
\begin{abstract}
\large Analysis of over 36 years 
of time series data from the 
NSO/AFRL/Sac Peak
K-line monitoring program 
elucidates five components of the variation
of the seven measured chromospheric parameters:
(a) the solar cycle (period $\sim$ 11 years),
(b) quasi-periodic variations (periods $\sim$ 100 days),
(c) a broad band stochastic process (wide range of periods),
(d) rotational modulation, and
(e) random observational errors, independent of (a)-(d).
Correlation and
power spectrum analyses
elucidate periodic and aperiodic 
variation of these parameters.
Time-frequency analysis illuminates 
periodic and quasi periodic signals,
details of frequency modulation due to differential
rotation,
and in particular elucidates the rather complex
harmonic structure (a) and (b) at time scales 
in the range $\sim$  0.1 - 10 years.
These results using only full-disk data 
suggest that similar
analyses will be useful at detecting and 
characterizing differential rotation in stars
from stellar light-curves such as those
being produced by NASA's Kepler observatory.
Component (c)
consists of variations over a range of timescales,
in the manner of a ${1 / f}$ random process
with a power-law slope index that varies in a systematic way.
A time-dependent Wilson-Bappu
effect appears to be present in the solar cycle
variations (a), but not in the more rapid
variations of the stochastic process (c).
Component (d) characterizes differential rotation 
of the active regions.  
Component (e) is of course not characteristic
of solar variability, but the fact that the observational
errors are quite small greatly facilitates the 
analysis of the other components.
The data analyzed in this paper can be found
at the National Solar Observatory web site 
 \verb+http://nsosp.nso.edu/cak_mon/+,
or by file transfer protocol 
at \verb+ftp://ftp.nso.edu/idl/cak.parameters+.
\end{abstract}


\normalsize
\section{The K-line Monitoring Program}
\label{introduction}

For nearly four decades the 
NSO/AFRL/Sac Peak
K-line monitoring program 
\citep{keil_worden}
has produced almost daily measurements
of seven parameters characterizing 
the chromospheric Ca II K-line 
integrated over the solar disk.
This program 
is aimed at characterizing 
chromospheric variability 
due to various processes 
and on various time-scales.

This is a good time to analyze these data
as they now cover more than three solar cycles 
(21, 22, and 23)
thus allowing comparison of two 
alternate cycles
as well as providing 
some preliminary information about the
beginning of cycle 24.
Beginning on November 20, 1976
and continuing to the present, 
the time series 
now cover more than three 11-year 
solar cycles or $1 {1 \over 2}$ Hale Cycles.
This paper describes analysis of the data up to
February 28, 2013.

The motivation for 
this survey and details of observational procedures
are given in
\citep{keil_worden,keil_1,white}.
Further documentation and data are 
available at \citep{nso_www}.
Table 1 of 
\citep{keil_1} 
describes the seven measured K-line parameters.
The order  listed below and the boldface tokens are as 
they appear in the data file posted at
\verb+http://nsosp.nso.edu/data/cak_mon.html+.

\begin{description}

\item {1.\emdx}: Emission Index, equivalent width in 1 \AA \ band centered
   on the line profile
   
\item {2.{\bf VIORED}  }: {$I(K_{2V}) / I(K_{2R}) = [ I(K_{2V}) - I(K_{3} ) ]/
                  [  I(K_{2R}) - I(K_{3} ) ]$}, ratio of blue to red emission maxima.
\item {3.{\bf K2VK3} }:$ I(K_{2V})/ I(K_{3})$, strength of blue wing relative to  $K_{3}$

\item {${\bf 4. DELK1}: \lambda(K_{2R}) - \lambda(K_{2V})$}, separation of the two emission maxima

\item {$ {\bf 5. DELK2}: \lambda( K_{1R} ) - \lambda(K_{1V})$}, separation of the two emission minima

\item {${\bf 6. DELWB}: \mbox{Wilson-Babbu  \ parameter}$}, 
separation of outer edges of emission maxima

\item {${\bf 7. K3}: K_{3}$}, intensity in the core of the line

\end{description}
\noindent
Further description of the parameters is as follows (quoted 
with reordering from the NSO web site):
\begin{quotation}
Several K-line parameters, including the emission index and various measures of asymmetry, are abstracted from the calibrated line profiles and stored on the NSO ftp site. These parameters include: 
(1) the Ca K emission index 
 which is defined as the equivalent width 
 of a 1 angstrom band centered on the K line core, 
(2) the line asymmetry which is the ratio of 
the blue and red K2 emission maxima (K2V/K2R),
(3) the relative strength of the blue K2 emission peak 
with respect to the K3 intensity (K2V/K3), 
(4) the separation of the two emission maxima (K2V-K2R), 
(5) the separation of the blue and red K1 minima (K1V-K1R), 
(6) the Wilson-Bappu parameter which is the width measured between the outer edges of the K2 emission peaks, and
(7) the K3 intensity (the core intensity).
\end{quotation}
The schematic
line profiles in
\citep{keil_worden} and in  Fig. 1 of \citep{donahue_keil}
illustrate these definitions.
Note that (1) and (7) are line intensities, 
expressed as an equivalent width and 
a percentage of the continuum, respectively;
(2) and (3) are intensity ratios, and
(4), (5) and (6) are 
wavelength separations of line features in Angstroms.

\begin{figure}[htb]
    \includegraphics[width=7in,height=3in]{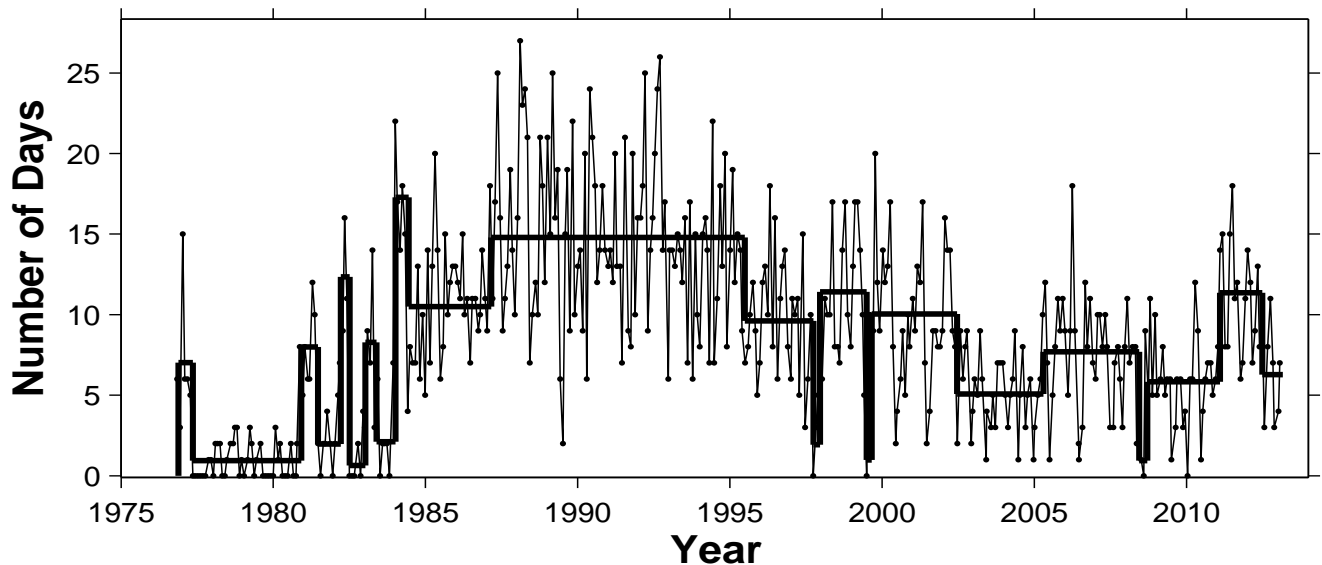} 
    \setlength{\baselineskip}{6pt}
    \caption{\baselineskip 16pt The points joined by
   thin lines are the number of days observed per 30 day interval.
   Thick solid lines indicate the optimal 
   piecewise constant representation of the
   observation rate.}
  \label{fig_obs_hist}
\end{figure}

Figure \ref{fig_obs_hist} shows the number of 
days on which observations have been obtained
during 30 day intervals. It is an update of
Figure 1 of \citep{keil_1}
in the same
same format 
but with slightly different interval boundaries.
If the observation times are 
independent random variables
with a changing rate (also known as
a variable-rate Poisson process)
the thick lines define 
the best step-function 
representation of the variation of the event rate,
obtained using the Bayesian Blocks algorithm
\citep{jackson,scargle_vi}.
\noindent
The mean interval between samples
was 3.39 days, and the median
interval exactly 1 day.

We here report exploratory analysis of these time
series data, aimed at
characterizing the variability of the
individual parameters in a number of ways, and to
investigate possible relationships between them. 
For background the reader may consult
the review paper \citep{hall} on stellar chromosphere
activity.
The book \citep{karel}
provides excellent overview of 
the relevant solar physics and stellar physics.
The paper \citep{livingston} details analysis
of McMath Solar Telescope data similar
to those described here.


The following sections describe 
time domain, correlation, 
power spectrum, and time-frequency 
analyses carried out on these data.
The emission index $\bf EM$
and core intensity $\bf K_{3}$
are emphasized, 
because 
these two closely related parameters
vary in quite similar ways and
seem to be the most straightforward diagnostics
of chromospheric activity.
No data preprocessing beyond
that described in \citep{keil_worden}
was applied,
other than the removal of a few outliers.

\section{The Time Series}
\label{time_series}

\begin{figure}[htb]
    \includegraphics[width=7in,height=8in]{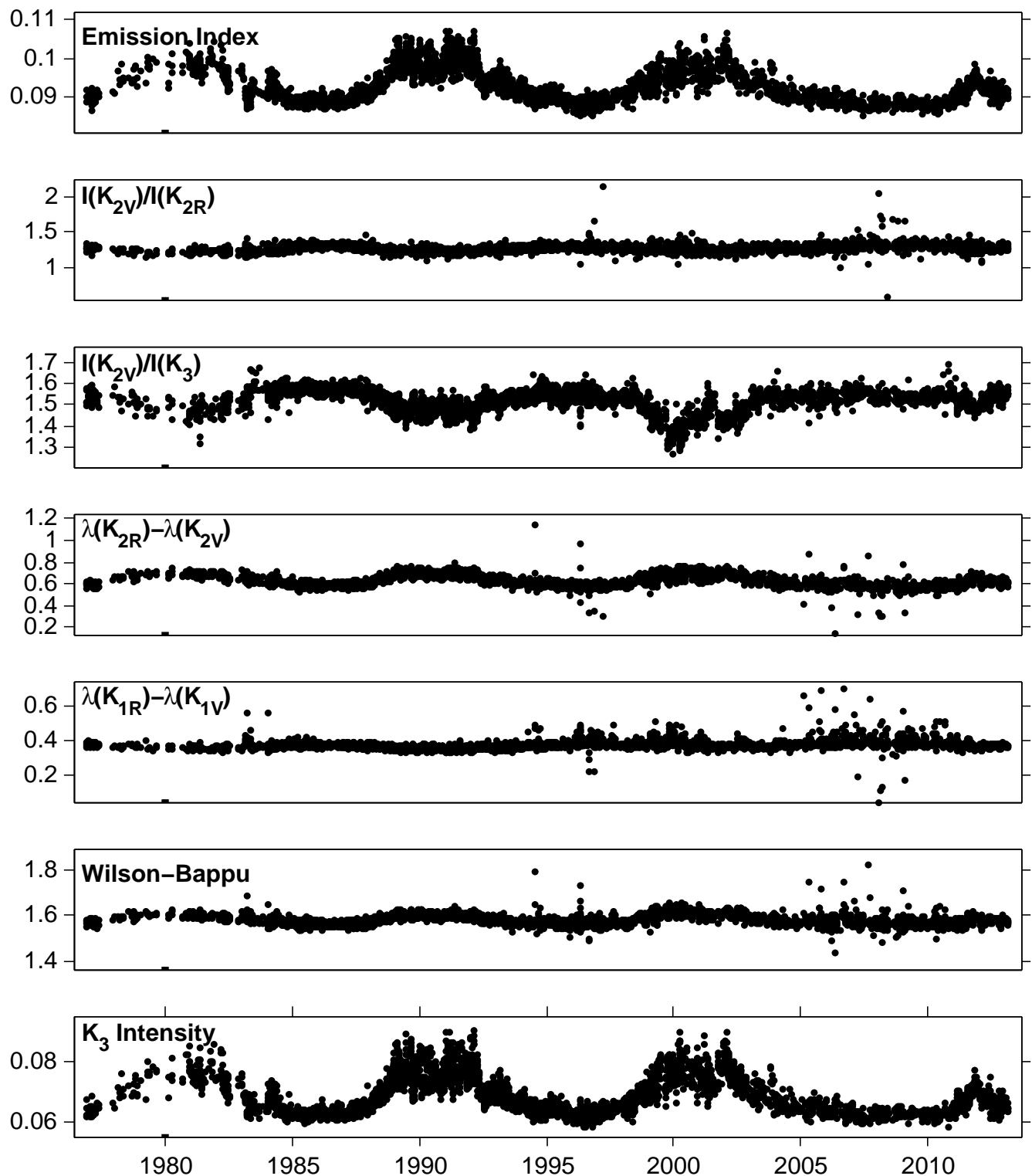} 
   \caption{\baselineskip 16pt Ca II K-line time series data, 
   with a few outliers removed.
   Just above the date 1980  
   is a small bar representing the upper limit
   on the average 
   error variance determined directly from the 
   data as described in \S \ref{auto_correlation}.
   }
\label{fig_all}
\end{figure}

Figure \ref{fig_all} presents these 3905 
observations in the
same format as Figure 2 of \citep{keil_1},
with the exception that the order is the same
as listed above, and a few outlying points
presumed to be erroneous have been 
replaced with linear interpolations.
In addition an estimate of the $1\-\sigma$ 
observational error variance is plotted as a 
small vertical bar near the bottom of each 
panel just above the date 1980.
These error bars are determined 
from an analysis of the auto-correlation
function of the time series data, as
described in \S \ref{auto_correlation}.
Note that these errors are quite small;
the majority of even the apparently 
random variation is real and
not due to observational errors.
\noindent
\clearpage

\begin{figure}[htb]
    \includegraphics[width=7in,height=8in]{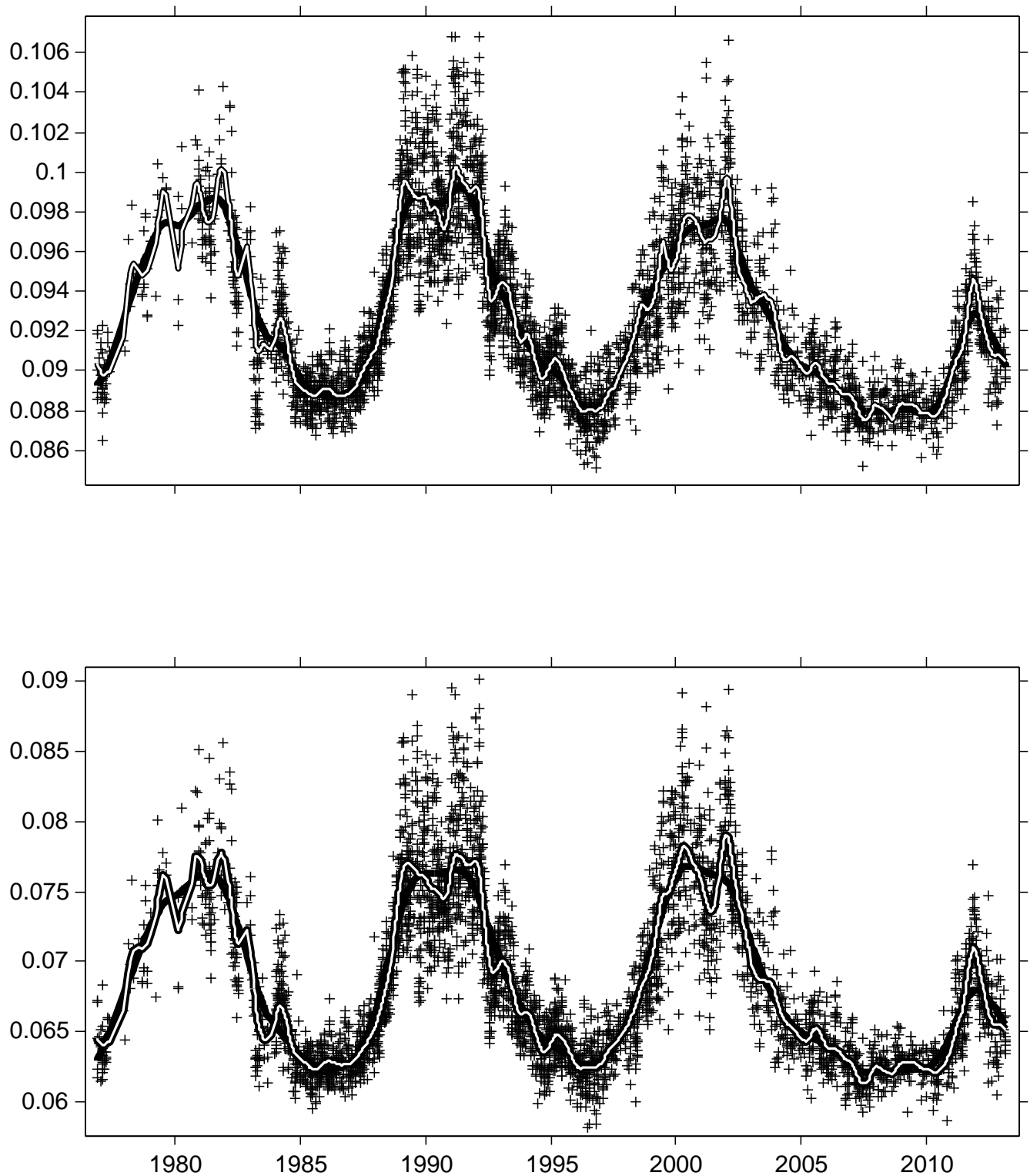} 
   \caption{\baselineskip 16pt Time series of  \emdx \ (top)
   and \kthree \ (bottom), with spline fits
   computed with two different degrees of smoothing:
   greater smoothing as the dark line, less smoothing
   as the white-filled line.}
   \label{fig_lc_17}
\end{figure}

Figure \ref{fig_lc_17} shows 
and enlarged version of the 
two intensity time series,
\emdx \ and \kthree.
The lines are fits to the data 
using the MatLab spline function \verb+spaps+.
The resulting smoothing has the effect of removing 
or reducing the shorter time-scale components, 
thus elucidating time-scales longer than a fraction of a year. 
Shown are fits with two different values for the 
spline error tolerance parameter.  
Roughly speaking the lesser smoothing reveals the
solar cycle 
and the somewhat faster quasi-periodic variations to be 
discussed below, 
while the greater smoothing mostly removes the latter
and emphasizes the former.

\clearpage
\noindent

\begin{figure}[htb]
    \includegraphics[width=7in,height=4in]{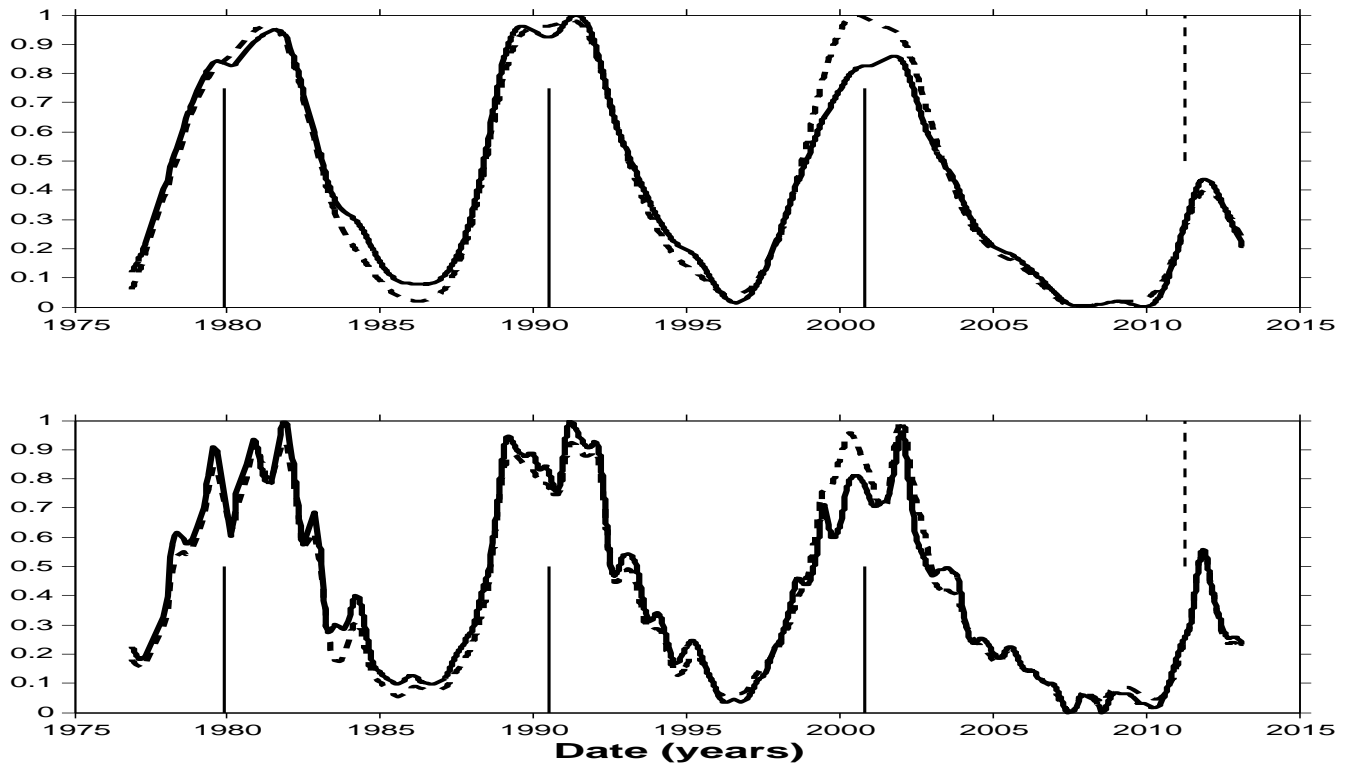} 
   \caption{\baselineskip 16pt Direct comparison of the two variables
   ${\bf EM}$ (solid lines) and ${\bf K_{3}}$ (dashed lines).
   Shown are spline fits to the raw data with outliners removed: 
   heavier smoothing in the top panel and 
   less smoothing in the bottom one. 
   }
   \label{fig_lc_comp_17}
\end{figure}

Figure \ref{fig_lc_comp_17} 
makes a side-by-side comparison of 
the two variables
\emdx \ and \kthree,
with both degrees of smoothing.
Figures \ref{fig_lc_17} and 
\ref{fig_lc_comp_17} between them
make two points:
(1) these two intensity variables,
under either of the adopted smoothing choices,
have very similar behavior;
and (2) the more complex structure corresponding
to the smaller degree of smoothing,
while not identical, 
is similar over the three cycles. 
This first point is 
not unexpected because
these variables measure similar  
aspects of the central depth of the K-line.
The second point is perhaps surprising, as 
it suggests that the solar cycle 
as seen in chromospheric activity 
is repeatably more complex 
than a series of 
simple monotonic
rises to maximum followed
by declines to minimum.
\noindent
We regard the repeatability of the irregular 
structures in the plots in the bottom panel 
as evidence that they more correctly represent the true 
behaviors of these parameters over the solar cycle
than do the smoother ones.
(Note especially cycles 21 and 23:
three sharp peaks near solar maximum,
with similar peaks on both the rising and falling
parts of each cycle.) 

As will be detailed in this 
and the following three sections,
there are four types of variability, 
plus observational errors, present 
in all of the time series:
(a) a periodic trend obviously tied to 
the 11-year solar cycle,
(b) quasi-periodic signals an order of magnitude faster than (a),
(c) random flicker noise, 
(d) a periodic signal at or about the solar surface rotation frequency,
and (e) the inevitable errors of observation.  
\begin{figure}[htb]\label{lc_rotation}
    \includegraphics[width=7in,height=3.5in]{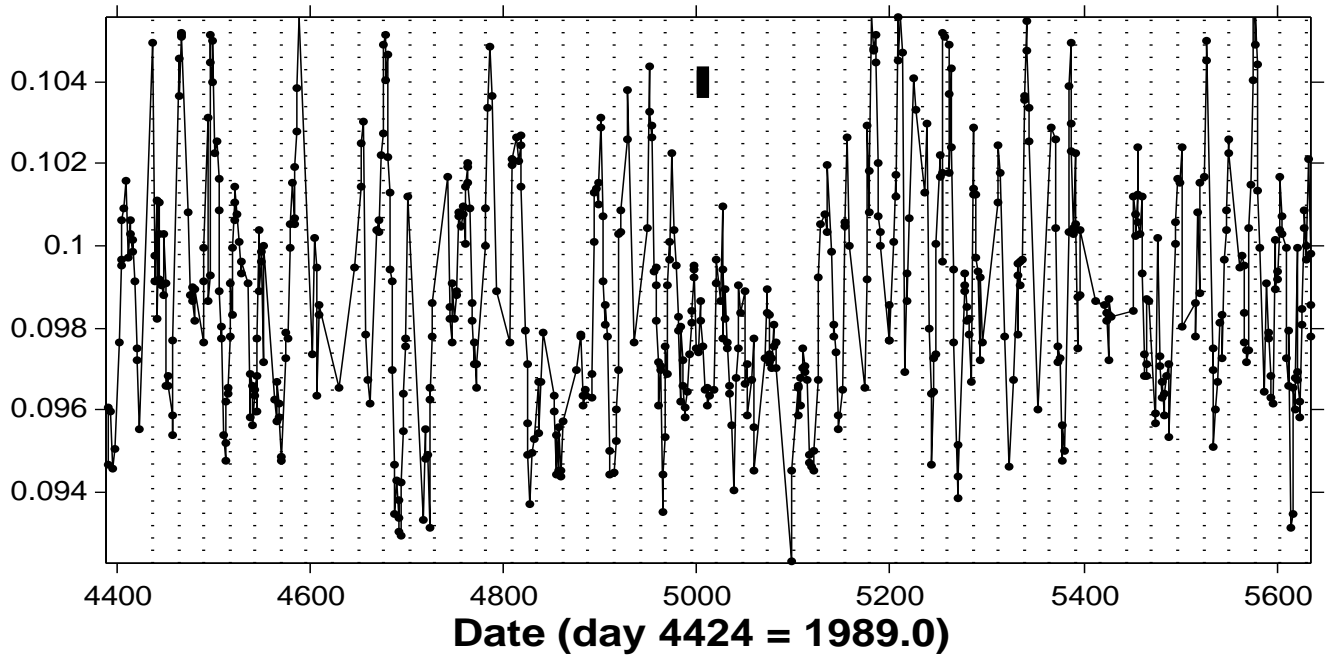} 
   \caption{\baselineskip 16pt A short segment of the \emdx \ light-curve
   near the peak of cycle 22.  The vertical dotted lines are separated
   by the period corresponding to the peak of the power
   spectrum of the data in this 3.4-year long interval.  
   (Accordingly this period, 26.49 days, 
   has meaning only for this limited time window.)    
   The solid bar
   in the middle/top part of the figure 
   is the $1\sigma$
   observational error variance determined in \S \ref{auto_correlation}. }
   \end{figure}
\noindent
The first two of these are the 
relatively smooth variations just
discussed.
The third and forth 
are difficult to distinguish from each other
visually in light curves. 
However the major part of the variability in 
the magnified plot in Figure 5 
is rotational modulation.
While there is not a precise one-to-one correspondence
between peaks and the fiducial lines at the solar 
rotation cadence, 
the presence of a periodicity
with an amplitude well above the observational errors
is strongly suggested.
In \S\ref{psa}  and \S\ref{tf}
all of these variation components 
(a)-(e) are separated from each other
using power spectrum and time-frequency analysis
of the residuals 
obtained by subtracting a 
smoothed fit from the raw data.
While this phenomenological separation 
may not mean that there really are 
four independent components, 
all of them are clearly real and originate from chromospheric 
activity, or a modulation thereof in case (d).
The observational errors are small,
as demonstrated in \S \ref{auto_correlation}.
In addition the details of the
variation of \emdx \ and \kthree \
are much the same (see Figure \ref{fig_lc_comp_17}), 
which would not be the case if observational errors
were significantly large.
It is difficult  \emph{a priori} to 
rigorously identify the physical processes
underlying these components, but the 
properties listed in 
Table \ref{table:variability}
argue for distinct physical origins
of the components.
\begin{table}[ht]
\caption{\baselineskip 16pt Five Modes of Variability} 
\vskip 0.1in
\centering 
\begin{tabular}{l l l l } 
\hline\hline 
\  & Amplitude & Time-Scale & Nature \\ [0.5ex]
\hline 
(a) Solar cycle & large & long ($\approx$11 years) & deterministic \\ 
(b) Rieger-type periods & small & medium ( $\approx$ 100 days)  & quasi-periodic \\ 
(c) Flicker noise & small & large range  & random \\ 
(d) Rotation modulation & medium & short (27 days)  & periodic  \\ 
(e) Observation errors & small & instantaneous  & random \\  [1ex] 
\hline 
\end{tabular}
\label{table:variability} 
\end{table}
\vskip 0.01in
\noindent
A positive amplitude-variance correlation is
clearly evident in the \emdx \ and \kthree \ time series,
and less prominently the others: 
variance large  near the peaks and small near the
valleys.
The plot of the residuals from a smooth fit
in Figure 6 makes this effect even more obvious.
\begin{figure}[htb]
   \label{em_res}
    \includegraphics[width=7in,height=3in]
    {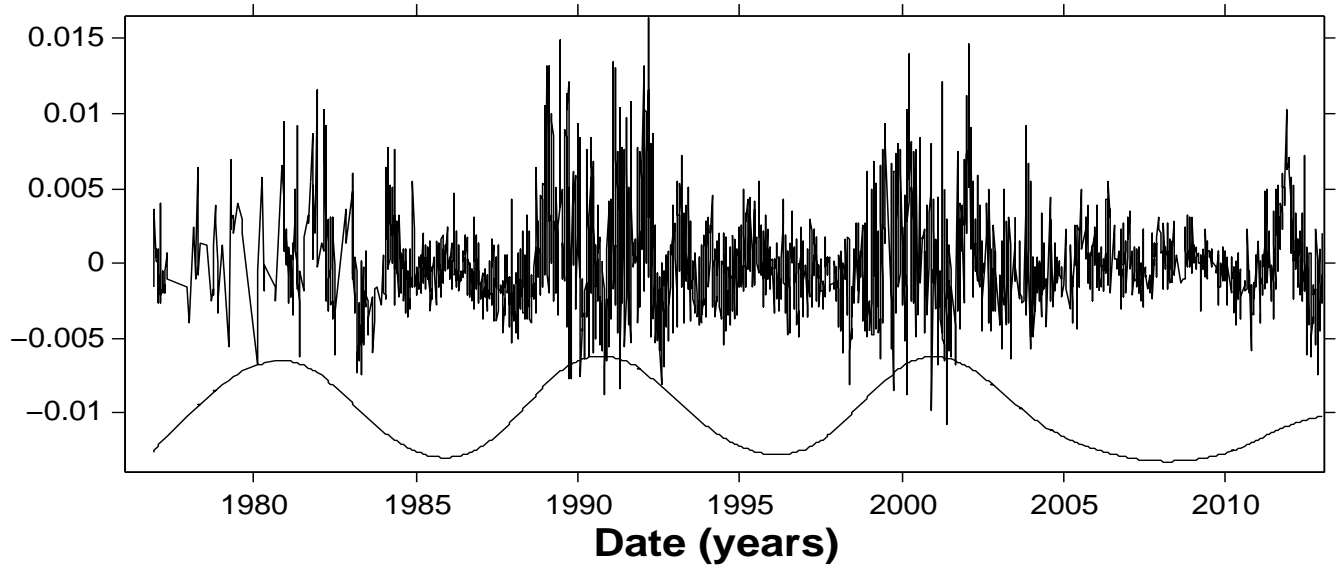} 
   \caption{\baselineskip 16pt Residuals of the \emdx \ data from
   the adopted smooth fit; this fit is plotted 
   at the bottom
   with an arbitrary offset and 
   scaled down by a factor of 2 realtive to the
   residuals.}
\end{figure}
Such correlations are expected,
in view of the 
large contribution to the variance from
rotationally modulated chromospheric activity
(\emph{cf.} Fig. 5)
closely following the solar cycle.
\clearpage

\begin{figure}[htb]
    \includegraphics[width=7in,height=3in]
    {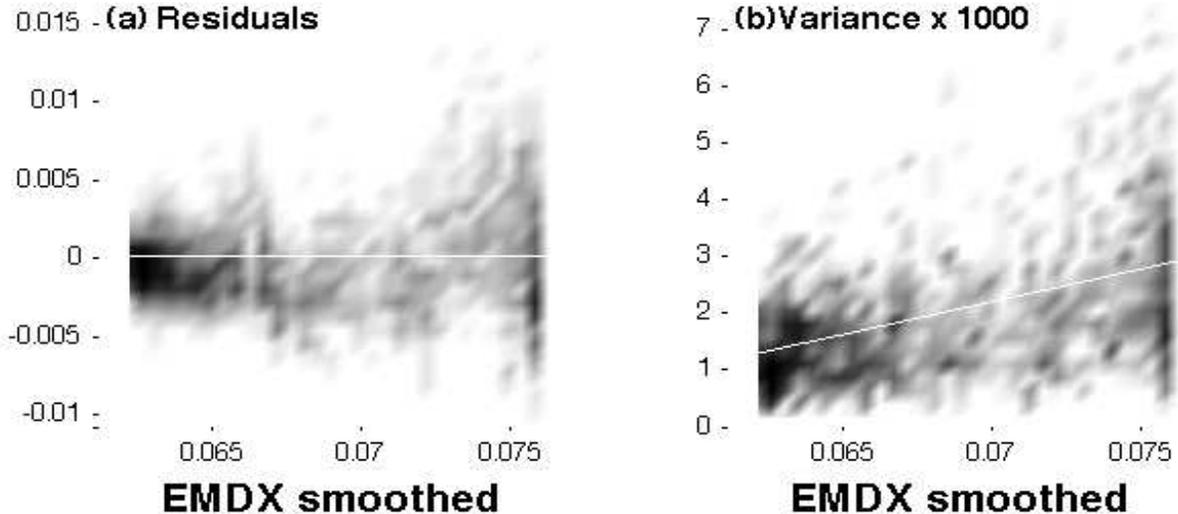} 
   \caption{\baselineskip 16pt Scatter 
   plots of  (a) \emdx \ residuals and (b) 
   variance of these residuals,
    vs. the smoothed \emdx \  parameter 
    are binned in a 32x32 
    grid of 2D bins and displayed as grey scale images.
    The superimposed white lines are (a) zero residual
    and (b) a least squares regression line.
   }
   \label{res_vs_smo}
\end{figure}
Figure \ref{res_vs_smo} is another way
to visualize the relationship between the
random and smoothed \emdx.
By construction the residuals average to zero, as in (a).
The increase of the variance with amplitude
is explicit in (b) and supported by 
the increase of the range of the residuals 
with emission in (a).

The following Table \ref{stat} presents summary 
statistics for the seven variables.
Except for the error estimates 
described in \S \ref{auto_correlation}
these were all computed in straightforward 
ways directly from the time series data.
The first three rows (after one
specifying the units for the quantity) 
contain 
the mean, range, and standard deviation 
computed directly from the raw observations
with outliers removed.
Row four is the standard deviation of the
residuals from the adopted smooth fit to the 
relevant time series.  
Row five gives the estimated 
RMS observational errors described
in the next section; these should 
be taken as upper limits
for reasons described there.
Row six is the relative error obtained by
dividing row five by row two.
Row 7 is the index $\alpha$ in power-law fits 
($P( f ) \sim f^{\alpha}$) to the power spectra, 
described in \S \ref{psa}.
\begin{table}[htdp]
\caption{\baselineskip 16pt Statistics of the K-Line Time Series}
\vskip .4cm
\small
\begin{tabular}{ || c | c | c | c | c | c | c |  c | r| c}
\hline
\\
&   EMDX  &  VIORED &  K2VK3  & DELK1 & DELK2  &    DELWB  &  K3 \\
  \\
\hline
\\
Units   & Eq W   & $\Delta$ intensity ratio  &  intensity ratio  &  $\Delta$\AA  &  $\Delta$\AA   & $\Delta$\AA  & intensity  \\
\\
\hline
\\
Mean   & 0.0929   & 1.2700  &  1.5146  &  0.6304  &  0.3748   & 1.5832  &  0.0687 \\
\\
Range &  0.0129 &   0.2256   & 0.2648  &  0.1862  &  0.0918  &  0.1031  &  0.0211\\
\\
$\sigma$ & 0.0043  &  0.0444  &  0.0549  &  0.0508 &   0.0204  &  0.0212 &   0.0062 \\
\\
Residual $\sigma$  & 0.0018  &   0.0321  &  0.0250  &  0.0229  &  0.0144  &  0.0116  &  0.0024 \\
\\
Error $\le$ & 0.0005   & 0.0303   & 0.0133  &  0.0180 &   0.0114  &  0.0108  &  0.0006 \\
\\
Error / Range &  0.0423  &  0.1343  &  0.0501  &  0.0967  &  0.1241  &  0.1052  &  0.0290\\
\\
$\alpha$ &  -0.303  &  0.004  &  -0.175 &  -0.108  &  -0.114  &  -0.009  &  -0.238\\
\\
\hline
\end{tabular}
\label{stat}
\end{table}%
\normalsize

\section{Autocorrelation Analysis}
\label{auto_correlation}

An autocorrelation function contains information 
about the memory of the underlying process, 
be it random or deterministic.
This function elucidates connections between 
the quantity at different times;
specifically the autocorrelation function $\rho(\tau)$
characterizes the joint variability at times $t$ 
and $t+\tau$ averaged over $t$.
(In the next section we will also use the
autocorrelation as a handy way to compute
power spectra and time-frequency distributions.)

\begin{figure}[htb]
    \includegraphics[width=7in,height=4in]
    {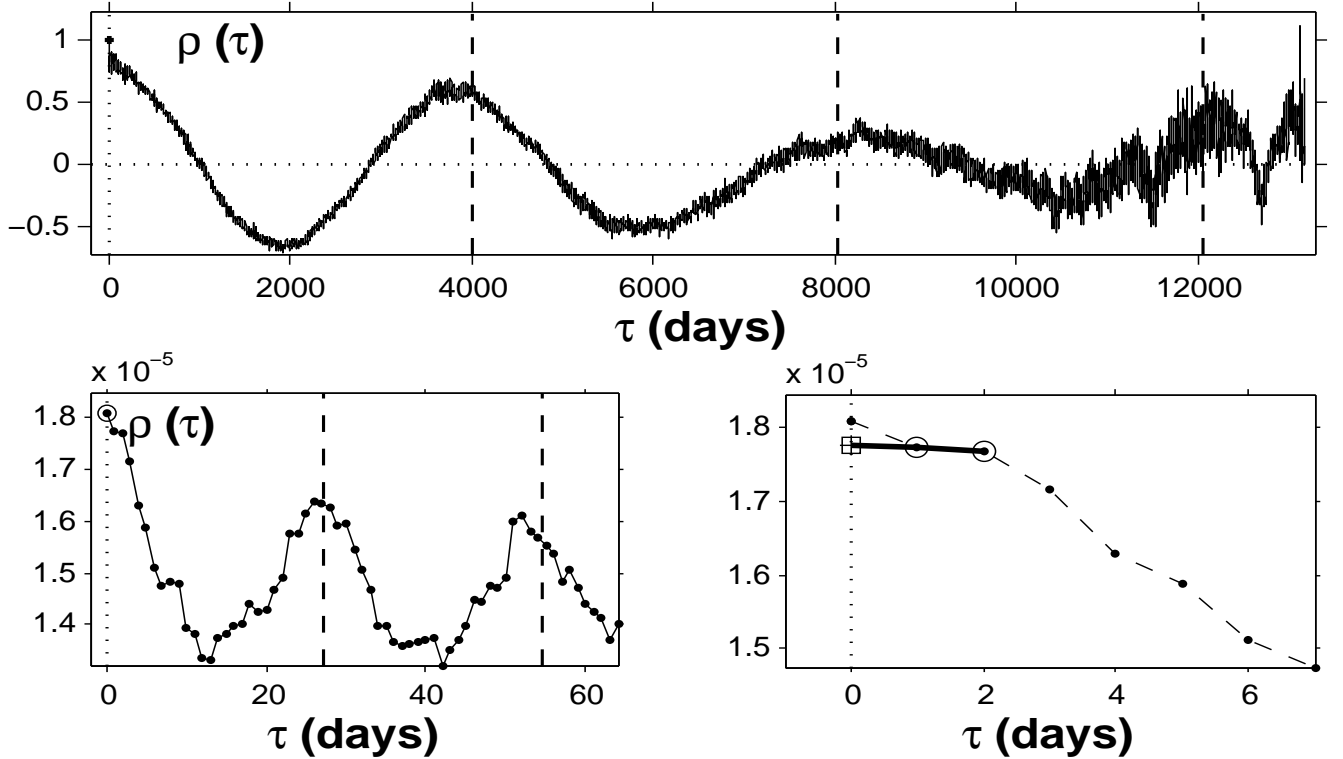} 
   \caption{\baselineskip 16pt Edelson and Krolik 
   correlation function of \emdx \ time series, with
    zero lag indicated by vertical dotted line.
   Top: Autocorrelation (normalized at zero lag) over the full range of lags.   
Bottom: unnormalized autocovariance at lags up to 
64 days (left) and 7 days (right). 
Lags of multiples of 11 years (top) and
the Carrington period of 27.2753 days (bottom)
are shown as dashed vertical lines.
}
   \label{em_acf_3}
\end{figure}

The panels of Figure \ref{em_acf_3} exhibit the 
rather complex multi-scale behavior of the 
auto-correlation function for \emdx \ --
computed using the Edelson and Krolik 
algorithm \citep{edelson} as described in Appendix 2 --
emphasizing three important time scales.
The first panel shows the \emph{autocorrelation function}
(normalized to unity at zero lag) extending 
to the maximum lag, namely 
the 12671 day length of the time series,
thus emphasizing time scales $\approx$  
the solar cycle.
The bottom two panels plot the unnormalized
 \emph{autocovariance function} (different
by only a constant factor, 
and indicating actual variances)
covering: lags in the range of the
surface rotation period, and 
the smallest times scales corresponding to 
the one-day sampling of the raw data, respectively.

The overall behavior of 
the autocorrelation
is dominated by
variability at the frequencies of the solar cycle
and the surface rotation.
In the top panel much of the scatter about what would otherwise
be a smooth function 
is due to a combination 
of several of the variability modes
listed in Table 1: the stochastic signal (c),
the rotational modulation (d), with a minor 
contribution of the errors (e).
The increased scatter 
for large lags (only shown in the top panel)
is simply due to the fact that 
for lags comparable to the length of
the observation interval many fewer data
points contribute to the average 
in Equation (\ref{eandk}) of Appendix 2.

In the bottom right panel of Fig. \ref{em_acf_3} 
as one considers smaller and smaller lags 
the autocorrelation levels out somewhat 
at two days and one day, and 
the value at zero lag is notably higher than this level.
This offset,
also visible in the other panels,
reflects the variance of the observational
errors and provides a way to 
estimate the average observational error in the data.
The auto-correlation function
at zero lag is the sum of two contributions:
the observational error variance and 
the true variance of the source.
At any other lag the 
errors average to zero
as long as they are uncorrelated.
These remarks yield a procedure for
estimating the size of 
the average observational error 
by attributing it to the excess contained 
in the zero-lag spike.  
Assuming the true autocorrelation
function is reasonably smooth, 
the difference between 
an extrapolation to zero lag
and the actual value $\rho( 0 )$
yields the variance 
corresponding to the observational errors.\footnote{In
essence we estimate 
$\sigma^{2}_{err} = \rho( 0 ) - \lim_{\tau \rightarrow 0}\rho( \tau )$.}
In the bottom-right panel of 
Figure \ref{em_acf_3} a linear fit to
the autocorrelation at the first two positive lags
(shown as circles) was extrapolated to 
the point contained in a square.  
For all seven parameters
this difference between the actual 
and extrapolated zero-lag values
is the error variance reported in Table \ref{stat}.

Actually these should be taken as upper 
limits on the true error variance.
Any solar variability confined to
time-scales shorter than the sampling interval
would make a contribution to the 
zero-lag spike that would be 
lost in our extrapolation procedure.
Although known turbulence and oscillations 
likely contribute in this way,
absent more quantitative information
on how smooth the true autocorrelation
function is on the diurnal time scale, 
the errors listed in
Table \ref{stat} are probably reasonably
good estimates.

\section{Power Spectrum Analysis}
\label{psa}

Rotation can produce a periodic
modulation of any solar time series.
This signal might be expected 
to be relatively weak 
in full-disk observations of chromospheric lines,
as discussed further in the next section, \S \ref{tf}.
Nevertheless
one goal of this work is to detect and
characterize any signatures of rotational
modulation in the K-line time series.  
This section demonstrates the 
rather strong rotational signal present in these
data and already remarked upon in
\S \ref{time_series} in connection
with Figure 5.  
Rotation yields 
peaks in the power spectrum
at the solar rotation frequency
and its harmonics.
Indeed even the more subtle effects
of differential rotation can be 
studied in considerable detail, as
shown in the following section.

\begin{figure}[htb]
    \includegraphics[width=7in,height=4in]
    {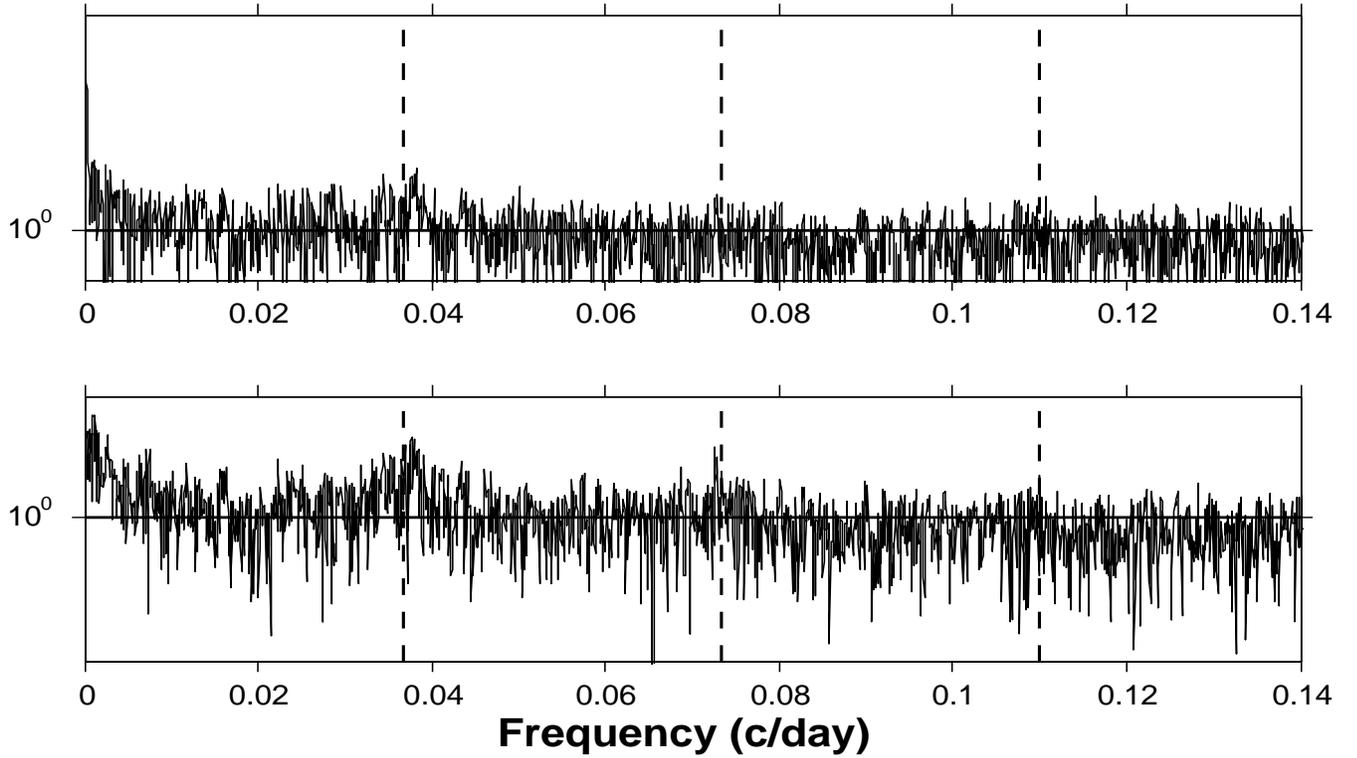} 
   \caption{\baselineskip 16pt \emdx \  power spectra, computed
   with the Lomb-Scargle periodogram, of:
   the raw \emdx \ data (top) and the residuals from smooth fit
   as shown in Figure \ref{em_res} (bottom). 
   Vertical dashed lines 
   denote the nominal 
   rotation frequency (corresponding to 
   the Carrington period of 27.2753 days)
    and harmonics.}
   \label{spec_1}
\end{figure}

Figure \ref{spec_1}
shows two power spectra for the \emdx \ time
series, both computed using the Lomb-Scargle
periodogram \citep{scargle_ii}. 
Power spectra computed
from  the Edelson and Krolik auto-correlation
function,
as mentioned in \S \ref{auto_correlation}
and detailed in Appendix 2,
are essentially identical to those shown here.
The comparison is between
the spectrum of the raw data (top) and that
of the residuals from the smooth fit (bottom).
The modulation at the rotation frequency 
and its harmonics is here
largely buried 
in the noise inherent in such unsmoothed power spectra.
It is slightly more prominent in the 
power spectrum of the residual data 
in bottom panel.

A feature of the power spectra 
of all seven parameters displayed in
Appendix 1 is a component at all frequencies, 
most easily seen as an approximately linear trend 
in the log-log plots.  That is to say
the power varies approximately as
$P(\nu) \approx \nu^{\alpha}$
with $\alpha$ taking on
negative values 
between 0 and -1.
Further $\alpha$ varies systematically
with time (lower-right panels in Appendix 1).
The most obvious feature of this variation is
a steepening of the power law (less high frequency
variability relative to that at lower frequencies) 
over a broad interval near the solar maximum
of cycle 22 (1990-1992).  
This behavior is perhaps related
to the fact that the overall activity in this cycle
was stronger than in the other cycles,
as can be seen in 
Figures \ref{fig_all}
and
\ref{fig_lc_17}, and
the top panel of 
Figure \ref{em_tf} for example.
Indeed the slightly less prominent 
high frequency background power
for cycle 22 is perceptible in this panel.

Raw power spectra are typically noisy and 
require smoothing in order to 
fully reveal their information content.
We do not pursue this avenue here, 
since an even more fruitful approach 
is detailed in the following section. 
\clearpage

\section{Time-Frequency Distributions}
\label{tf}

Rotation induces 
a harmonic variation of any signal 
from a localized region of the Sun's surface and
observed from a fixed direction.\footnote{Observed 
from the Earth 
the rotational modulation of such a region 
is approximately a truncated sinusoid, with
Fourier components at the frequency 
corresponding to relevant synodic period,
plus harmonics.}
In the case of spots and active regions
differential rotation 
modulates the observed period
as they experience 
the latitude drift tied to the solar cycle.
The summation of sources at various 
solar longitudes and latitudes
inherent to full disk observations 
smears out the power spectrum 
and dilutes the signature of differential rotation.
Nevertheless a surprising amount of 
information about the Sun's rotation 
is contained in the full-disk K-line time series,
as we shall now see.

The \emph{time-frequency distribution} 
is designed specifically 
to explore this sort of 
evolving harmonic structure.
In a nutshell 
this signal processing tool displays the time evolution of 
the power spectrum.
The excellent treatise  \citep{flandrin}
explains what can be learned 
with the basic tool and 
a number of its variants.
Here we use the simplest approach, namely 
computing power spectra of 
the data within a sequence of time windows 
covering sub-intervals of the full observation span.
Accordingly this tool is also called a
\emph{dynamic}
or \emph{sliding window} power spectrum.
The output is a three dimensional data structure -- 
power (z-axis) as a function of time and frequency
(x- and y-axes) -- that we here render as 2D
grayscale plots.

A slice of this plot parallel to the frequency axis 
contains the power spectrum (power vs. frequency)
at a specific time.
A slice parallel to the time axis
depicts the time dependence of power
at a specific frequency.
The same mathematics leading to 
the Heisenberg uncertainty principle
dictates that these slices'
resolutions
cannot both be made small independently:
good frequency resolution can be had only 
with relatively large time windows,
and  good time resolution requires 
short windows.\footnote{Simply decreasing the
time increments by which the window is moved does
not increase the time resolution. It is the
size of the window that fixes the smoothing 
in the time-domain.}
Any implementation of the
time-frequency distribution 
allows one to mediate this unavoidable 
resolution trade-off,
for example 
by adjusting the size of the window.
A few further details of the computation 
of time-frequency distributions are
given in Appendix 2.

Figure \ref{em_tf} shows time-frequency plots
for the emission index.  
Because the data are not evenly spaced,
all of the time-frequency distributions
shown here were computed using the 
interpolation-free techniques
described in Appendix 2.
The cross symbols 
near the top right corners indicate
the time and frequency resolution. The length of the
arms of the cross indicate the width of
the sliding window and the corresponding
fundamental frequency.
The solid, dashed, and dot-dashed vertical lines
mark the frequency range corresponding 
to the rotation rate as a function of 
solar latitude, using Equation (3) of
\citep{brown}:
\begin{equation}\label{brown}
{\Omega_{p} \over 2 \pi} = 
452 - 49 \mu^{2} - 84 \mu^4  \ \ \ \mbox{nHz} \ ,
\end{equation}
\noindent
labeled with the corresponding
latitudes (in the $0$ to $60$ degree
range normally inhabited by spots).
The dotted line labeled ``R'' is the frequency 
corresponding to the Rieger period, taken as 155 days.
\begin{figure}[htb]
    \includegraphics[width=7in,height=7in]
    {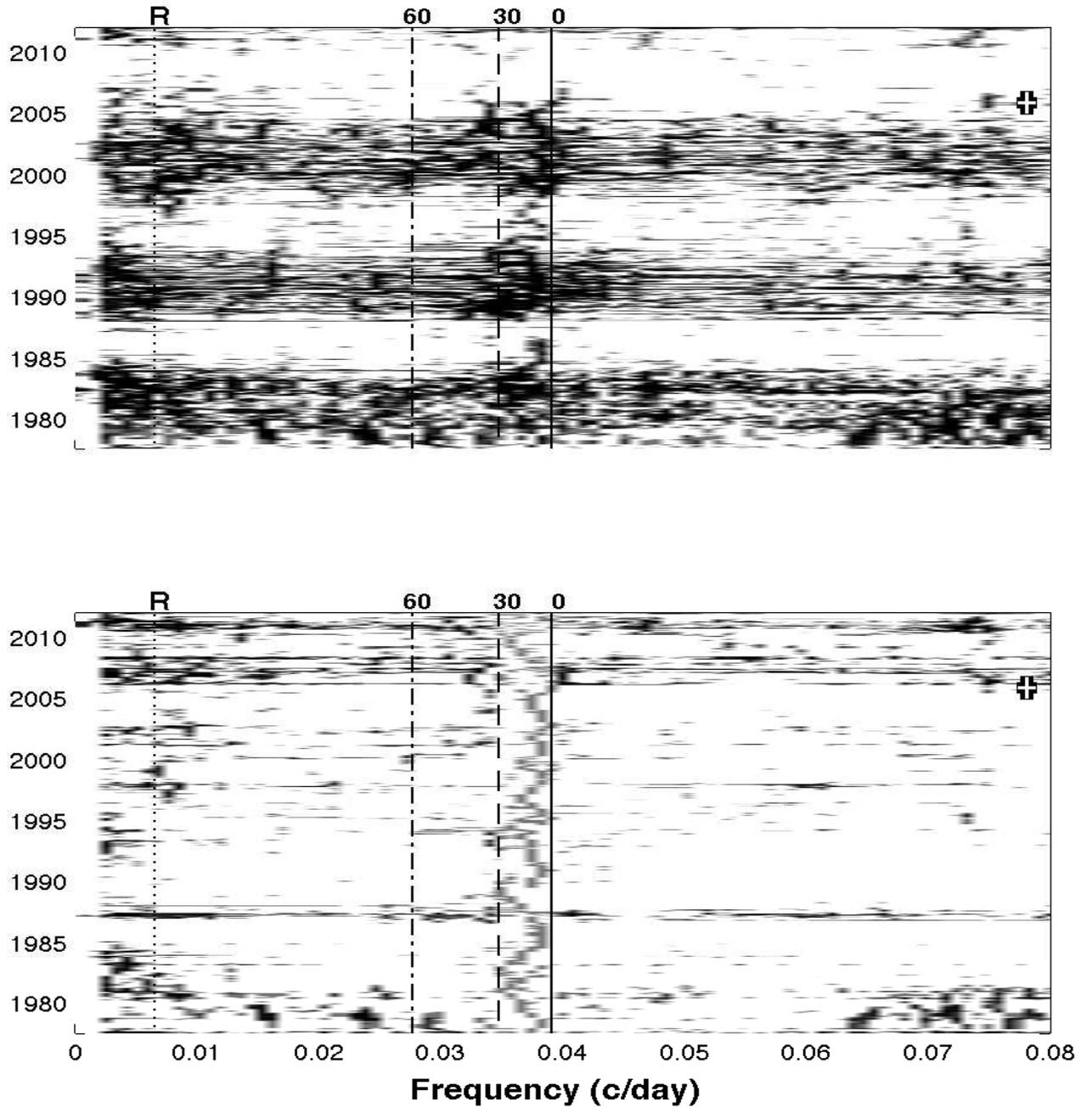}
   \caption{\baselineskip 16pt Time-Frequency distribution
   of \emdx \ residuals.
   Bottom, renormalized to make constant the 
   power between the frequencies corresponding 
   to 0 and 30 degrees, 
   bringing out behavior 
   during solar minimum that is otherwise lost. 
   Vertical 
   lines denote latitudes 0, 30, and 60 $\deg$
   based on Eq. (\ref{brown}).
   The dotted line at
   .0082 c/d roughly corresponds to quasi-periodicities
   discussed in \S \ref{psa}.
   Power 
   below .00176 c/day
   is divided by 10 to improve the display. }
   \label{em_tf}
\end{figure}
\clearpage

In the top panel of Figure  10 
the rotational features almost disappear
during solar minimum and are generally strongest 
in the middle cycle 22.
It is instructive to adjust for these effects 
by renormalizing each time slice of the
distribution.
Note the interesting behavior of the rotational signal 
in the bottom panel, which is 
a renormalized version of the top panel.
A relatively well-defined peak in power moves 
back and forth 
between approximately 0 degrees and 30 degrees
latitude and is present essentially all of the time,
not just during solar maximum as
one might have concluded from the top panel.

\begin{figure}[htb]
    \includegraphics[width=7in,height=4in]
    {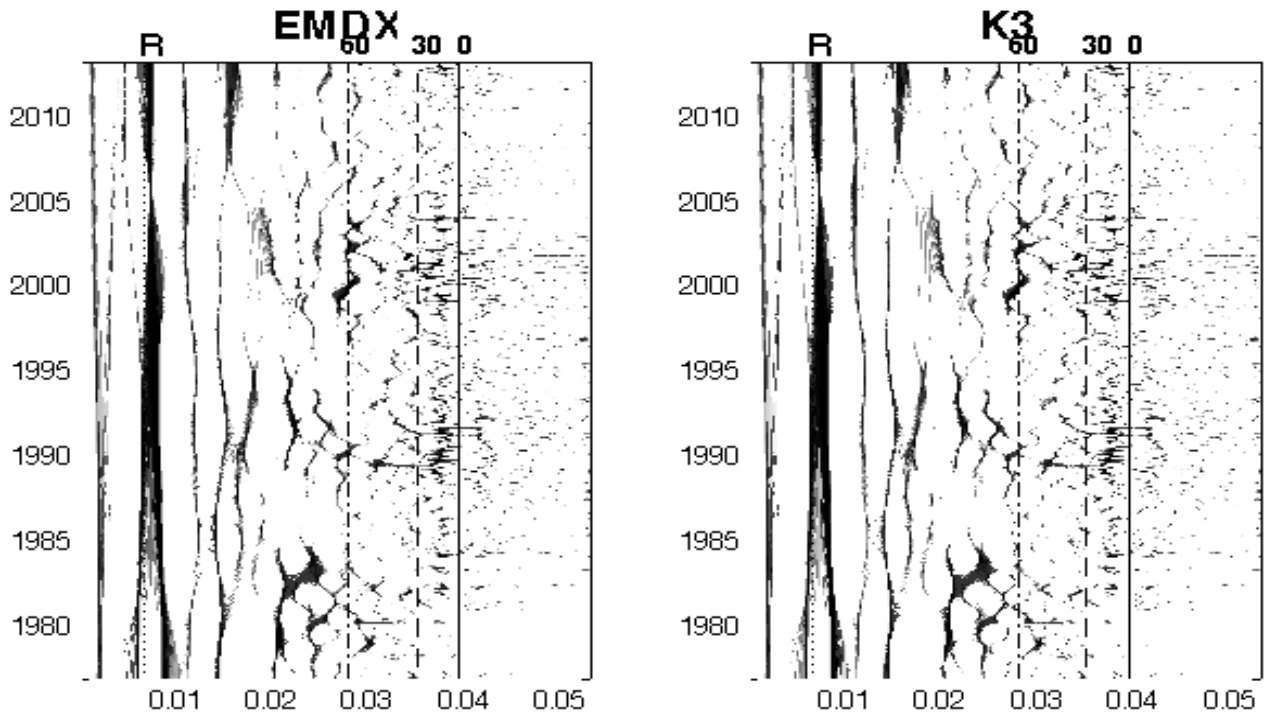} 
   \caption{\baselineskip 16pt  \emdx \ and \kthree \  time-frequency 
   distributions computed 
   with the synchrosqueezing algorithm
   of Brevdo and Daubechies.
   The number-of-voices
   parameter = 50. 
   The vertical lines represent the same nominal 
   frequencies bracketing differential rotation 
   as in Figure \ref{em_tf}. }
   \label{em_sync}
\end{figure}

The signal processing literature contains descriptions of
many other ways to estimate time-frequency distributions \citep{flandrin}.
One of the most recent ones, 
called \emph{synchrosqueezing},
represents the time series as 
``the superposition of a (reasonably) small number of components, 
well separated in the time-frequency plane, 
each of which can be viewed as approximately 
harmonic locally, with slowly varying amplitudes'' \citep{daubechies}.
Figure \ref{em_sync} shows the result of this
analysis for \emdx \ and \kthree \ 
interpolated to even spacing, using the  
MatLab tools in \citep{brevdo}.
The gray scale represents the power spectrum
of the variables, as a function of time and frequency.
There are broad similarities 
to the distributions in the previous figure,
and the clear differences in detail
can be understood in terms of the 
constraints imposed by the synchrosqueezing
method on the time-frequency 
atoms, as well as the fact that interpolation
to even spacing was necessary.  
The more fully non-parametric
sliding-window Fourier power spectrum 
spectrum yields a different representation of the underlying
time-frequency structure.
The synchrosqueezing algorithm
renders the differential rotation features
as more discrete 
than does the sliding-window approach;
the reverse is true of the quasi-periodic
signals with periods in the vicinity of 
0.002-0.015 cycles per day.

\section{Cross- Analysis}
\label{cross}
Each of the seven K-line parameters 
probes a different aspect of the chromosphere.
Therefore relations between the
corresponding time series
can elucidate physical processes
driving the underlying activity.
For example variability in two parameters
that is correlated, anti-correlated, or correlated
with time-lags can shed light on 
underlying dynamical processes.  
Of course causality cannot
be proven in this way, but relationships
consistent with physical models can be
elucidated.

\begin{figure}[htb]
    \includegraphics[width=7in,height=7in]
    {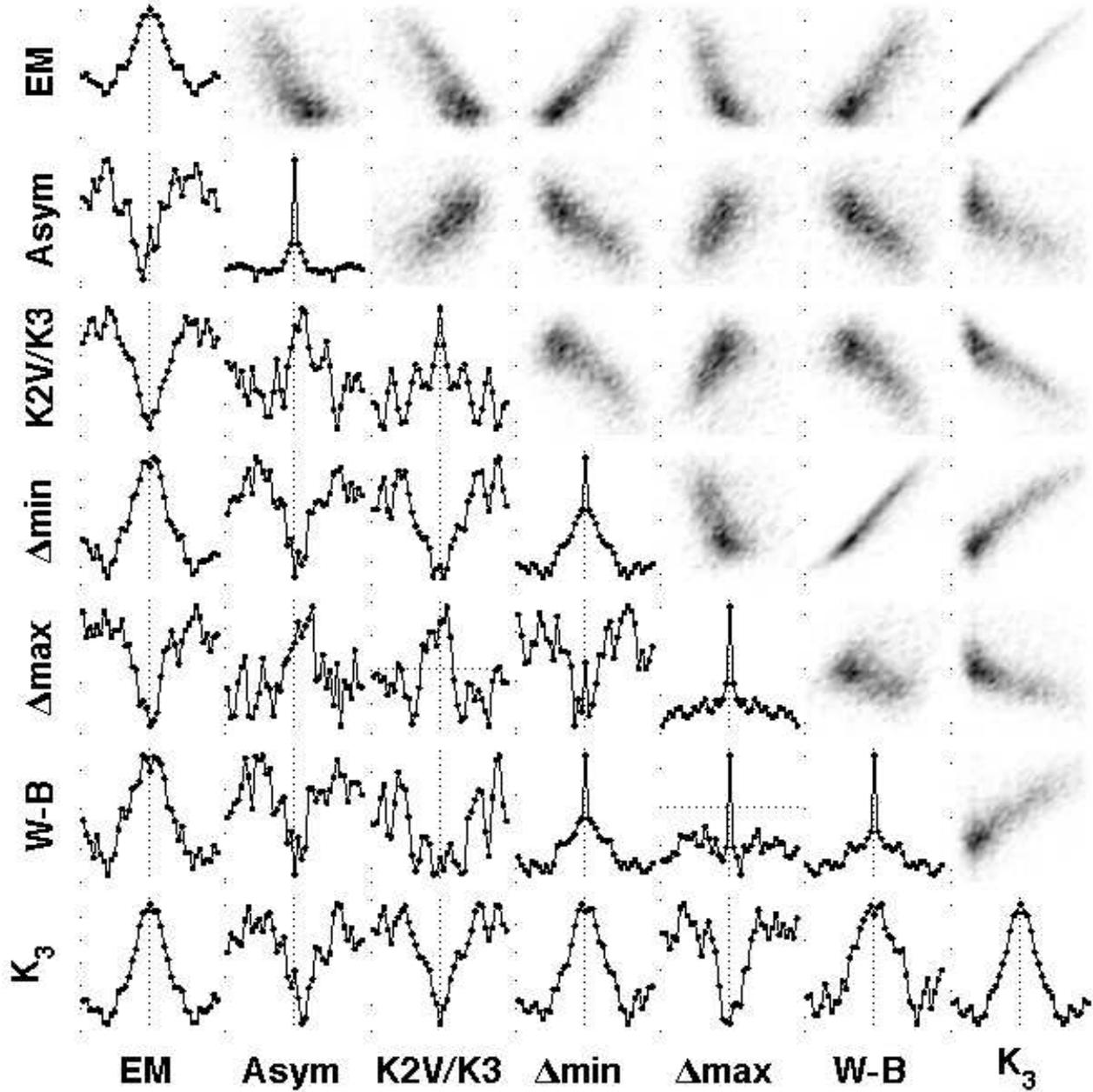} 
   \caption{\baselineskip 16pt Diagonal: Auto-correlations.
   Above Diagonal: Scatter plots.
   Below Diagonal: Cross correlations.
   All correlations are for $|\tau|$ $\le 20$ days. 
In computing the quantities
displayed in this figure
all seven parameters (with outliers removed) 
were standardized to zero mean and unit
variance.
    }
   \label{cross_all}
\end{figure}

Figure \ref{cross_all} depicts two types of
cross-analytic relationships for all pairs
of parameters.
The 21 scatter plots above the diagonal 
describe pairwise mutual dependence.
Below and on the diagonal are cross- 
and auto-correlation functions, respectively;
all were computed with the Edelson and Krolik
algorithm (\emph{cf.} \S \ref{auto_correlation} and
Appendix 2).

These types of displays 
are complementary ways of relating two variables.
Independence is a stronger statistical 
condition on two variables
than their being uncorrelated.
The former implies the latter, but not \emph{vice versa}.
Hence to the extent that scatter plots 
elucidate dependence they are more 
powerful statistically.
However they
depict only simultaneous relationships,
whereas cross-correlation functions
elucidate how the variables at two different
times are related.


\clearpage

\section{Wilson-Bappu Effect}
\label{wilson_bappu}

\begin{figure}[htb]
    \includegraphics[width=7in,height=5.5in]
    {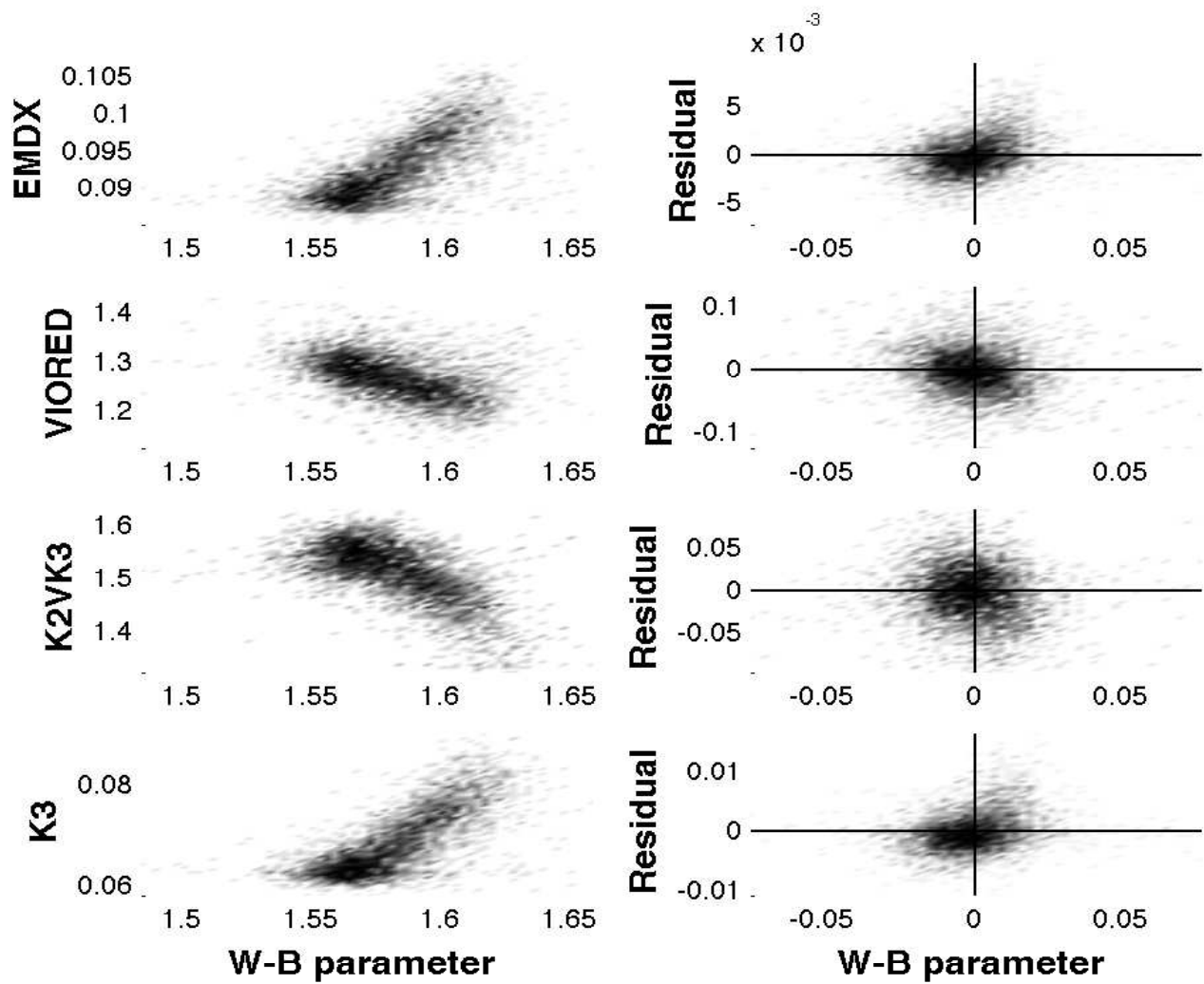} 
   \caption{\baselineskip 16pt Intensity variables vs. Wilson Bappu parameter. 
   Left-hand column: raw measurements (with outliers removed).  
   Right-hand column: residuals}
   \label{wb}
\end{figure}

In these data there is evidence for 
a Wilson-Bappu effect 
in the sense that the intensity parameters 
correlate with the width of the K-line.
Figure \ref{wb} contains scatter plots for
the four intensity parameters vs. the
Wilson-Bappu parameter, 
presented as $64$ by $64$ two-dimensional histograms
portrayed as greyscale plots.
The left-hand column contains scatter plots for the 
raw data (with outliers removed).
These correlations are presumably due to 
chromospheric processes 
tied to the variations in physical conditions
over the solar cycle.
The right-hand column shows 
the corresponding residuals from 
the smooth fits described in \S \ref{time_series},
which are essentially uncorrelated.
\clearpage

\section{Conclusions}
\label{conclusions}

This is basically an exploratory study of the
data collected in the K-line monitoring program.
The main goal is to follow the clues provided by 
various time series analysis methods
in the time-, lag-, frequency- and time-scale domains,
rather than to validate specific astrophysical hypotheses.

Affecting the interpretation of any results is their
significance in light of random observational errors
and systematic errors.  
Even though point-by-point errors
are not available,  in \S \ref{auto_correlation}
we have placed 
reasonable upper limits on the average
error variance.
In addition, it is difficult to construct and
display errors on 2D and especially 3D
functions such as time-frequency distrbutions.
We rely on comparison of the various functions 
to indicate the rough importance of observational
errors.  For example 
the time-frequency distributions for \emdx \ 
and \kthree \ for the most part show the
same beavior and therefore indicate that
such errors are not large.
The other parameters show signficantly
different behavior, but comparison of 
similar ones (such as the three wavelength
differences) gives a similar indication of
the signficance of the time-frequency structures.

In \S \ref{time_series} we presented evidence for a 
relatively complex structure to the solar cycle,
somewhat different from the standard  
concepts of the sunspot cycle
and other heliospheric phenomena 
discussed by a number of authors.
Apparently the K-line features are particularly 
sensitive to the changing physical 
conditions during the solar cycle.
A carefully chosen degree of smoothing
of the time series 
is essential to elucidating this structure. 

The basic notion of complexity 
in the solar cycle is not new, although
previous work has centered mainly on 
a relatively simple ``double peak'' concept  
in sunspot and other heliospheric indices
at the time of solar maximum.
For example, in yearly averages of 
the number of intense geomagnetic storms
\citep{gonzalez} describes time series behavior
that generally follows the solar cycle, but with a
double peak structure: `` ... one at the late
ascending phase of the cycle and another at the
early descending phase,'' with hints of even
more complex three-peaked structure 
for cycle 20.
In \citep{hathaway}
Figures 16 and 38 
show distinct double peak structures in
the smoothed International Sunspot Number,
and Figure 42 shows the same for the Polar
Magnetic Field Strength
as measured at the Wilcox Solar Observatory.
One presumes that the weakness of apparent 
double peaks in sunspot number
averaged over cycles 1 to 22 
depicted in that paper 
could be because of slight variations of the
times of the peaks as well as the degree of
smoothing applied to the individual curves.
Figure 9 in
\citep{domingo} depicts
a double peak and more complex 
structure in the time series for sunspot numbers,
10.7 cm. radio flux, a Ca K 1 $\AA$ full-disk index
from Kitt Peak, and a Mg II index.
The reference \citep{khramova} gives an
overview of structure in sunspot variability
on various time scales, referring to the
kind of structure noted here as ``quasi-biennial
oscillations.''
The complex time series structures shown in
the bottom panel of our Figure \ref{fig_lc_comp_17}
perhaps correspond to multiple toroidal field surges 
as discussed by 
\citep{georgieva}.

A somewhat related issue 
is the structure noted 
in the time frequency distributions,
possibly connected to the solar cycle
but at frequencies lower than 
those due to surface rotation.
There is discussion in the literature
of such frequencies, often in the context
of an early claim of a 154 day
periodicity in solar gamma-ray flares
\citep{rieger}, which was followed by
attempts to find similar periods in other phenomena.
\citep{sturrock} discusses an idea 
in which a more complex structure
consisting of multiples of a fundamental
period of approximately 25.5 days
underlies the Rieger periodicity; see also
\citep{bai,sturrock_and}.
\citep{hill} discusses a period of 151 
days in solar cosmic ray fluxes.
\citep{joshi} find a 123-day period in 
soft x-ray flux from the sun, and
\citep{lou} find a very similar period 
(and others) in solar coronal mass ejections.
The relevance of similar periodicities
occurring in other stars
\citep{massi} is unclear.

The K-line data as analyzed in the
time-frequency distributions in \S \ref{tf}
suggest the presence of some 
quasi-periodic behavior 
on similar time scales.
We found that peaks in a sine wave 
of suitably chosen period and phase matches
many of the peaks in the partially smoothed
\emdx \ time series.
A period of 122.4 days 
was obtained in a rough peak-fitting
procedure.
However, keeping in mind the 
degrees of freedom in the sinusoid,
the uncertainty in locating and timing 
the peaks in the data, 
and the matching of some peaks and not others,
this result does not prove the
existence of a pure harmonic signal.
Rather as indicated in the time-frequency
distributions there appears to be
quasiperiodic behavior in this frequency
range.

Another result of our analysis is 
the characterization of differential surface rotation,
using the time-frequency tool, as described in
\S \ref{tf} for the main intensity variables,
with displays for all of the variables in Appendix 1.
For discussions of differential
rotation estimated from time series from
Kepler and CoRoT see
\citep{frasca} and \citep{silva} respectively.

Here is a summary of 
our broad conclusions:
\begin{itemize}
\item The solar cycle variability [component (a) in Table \ref{table:variability}] 
of the K-line intensity consists of the well-known broad oscillation paralleling the 11-year sunspot cycle.

\item In addition there are quasi-periodic
oscillations that do not have constant periods or
amplitudes, but irregularly populate the time-frequency
domain in the neighborhood of periods of roughly 100 days.
It is not clear if these are modulations of the solar
cycle or a physical process independent of same.

\item The random variations
["Flicker noise"; component (c) in 
Table \ref{table:variability}] in all seven parameters 
have power spectra describable 
as ``${1 \over f}$ noise,'' meaning $P(\nu) \approx \nu^{\alpha} $.
The index $\alpha$ is always negative
and exhibits systematic variations over time 
apparently correlated with the general level of 
chromospheric activity.  Perhaps higher activity
is due to many individual independent fluctuations,
the summation of which effectively smoothes 
higher frequency variations as a purely statistical
effect. Or some other inherent feature of the 
randomness of the underlying physical process 
may suppress rapid dips from states of higher 
to lower activity.

\item Components (a) and (b) are, 
roughly speaking, independent of
each other, except that the variance of (b) is 
correlated with (a) [Figures 6 and 7].

\item A signature of differential surface rotation is 
captured by time-frequency analysis of especially the \emdx \
time series.
While this behavior roughly mirrors the general 
character the butterfly diagram for sunspots,
in detail it is distinctly different.
These differential rotational signatures 
of the K-line parameters
continue during solar minimum.

\end{itemize}
\noindent

These conclusions refer mainly to the
the parameters \emdx  \ and \kthree, but 
to some degree apply also to others 
of the measured parameters.


Acknowledgements:  We are grateful to
Alexander Kosovichev,
Kira Rehfeld and Luca Bertello
for helpful comments, 
and to Joe Bredekamp
and the NASA Applied Information Systems Research Program
for encouragement and support.
The observations used herein 
were obtained at the Evans Solar Facility 
of the National Solar Observatory. 
We are grateful for the assistance of NSO personnel, 
especially John Cornett, Timothy Henry and Lou Gilliam 
for observing and reduction of the raw data 
to produce the Ca II K-line data archive. 
The NSO is operated by the Association of Universities for Research in Astronomy, Inc. (AURA), for the National Science Foundation.

\newpage
\begin{center}
Appendix 1: Power Spectra and Time-Frequency Distributions
\end{center}

This appendix presents power spectrum analysis
of all of the measured variables.  
In each of the seven figures
the top two panels show the 
Edelson and Krolik based power spectra
of the residual time series,
plotted linearly (left) and doubly logarithmically (right).
A ${1 \over f}$ power spectrum 
corresponds to a straight line in 
the latter; a dashed line shows the corresponding 
least-squares fit
excluding the regions of the rotational peaks.
Vertical lines indicate 
frequencies corresponding to a 27.2753-day 
nominal surface rotation period
(the Carrington period) and its harmonics.
The vertical dashed line in the upper right
panel marks a fiducial frequency
corresponding to a period of one year.
The 11-year frequency is too small to
plot effectively here.

The lower-left panel shows the time-frequency distribution
obtained by computing the power spectrum
in a time-window slid along the time series.
These were normalized much as in the
bottom panel of Fig. 10, but over the broader 
frequency range 0.02-0.04 cycles/day.
The power spectra shown in the top two panels
were obtained by averaging this time-frequency distribution 
over the time coordinate.
Accordingly they are considerably smoother 
than would be obtained directly from the time series. 
The upper-right panel shows 
the full frequency range
extending to the Nyquist frequency (${1 \over 2}$ cycle/day),
but in both of the left-hand panels 
only a restricted range covering
the most interesting behavior
is plotted.


The bottom-right panel shows the
temporal 
variation of the slope of the power-law
fit to the power-spectrum -- that is the value of
$\alpha$ in a representation of the form 
\begin{equation}\label{power_law}
P( f ) = P_{0} f^{\alpha} \ .
\end{equation}
\noindent
Note: we adhere to
the convention that a process 
that even approximately 
satisfies this equation with
any value of $\alpha$ (almost always
negative) is called ``${1 \over f}$ noise''.
Comments on the systematic 
variation of this parameter are
contained in Section \ref{conclusions}.

\begin{figure}[htb]
    \includegraphics[width=7in,height=8in]{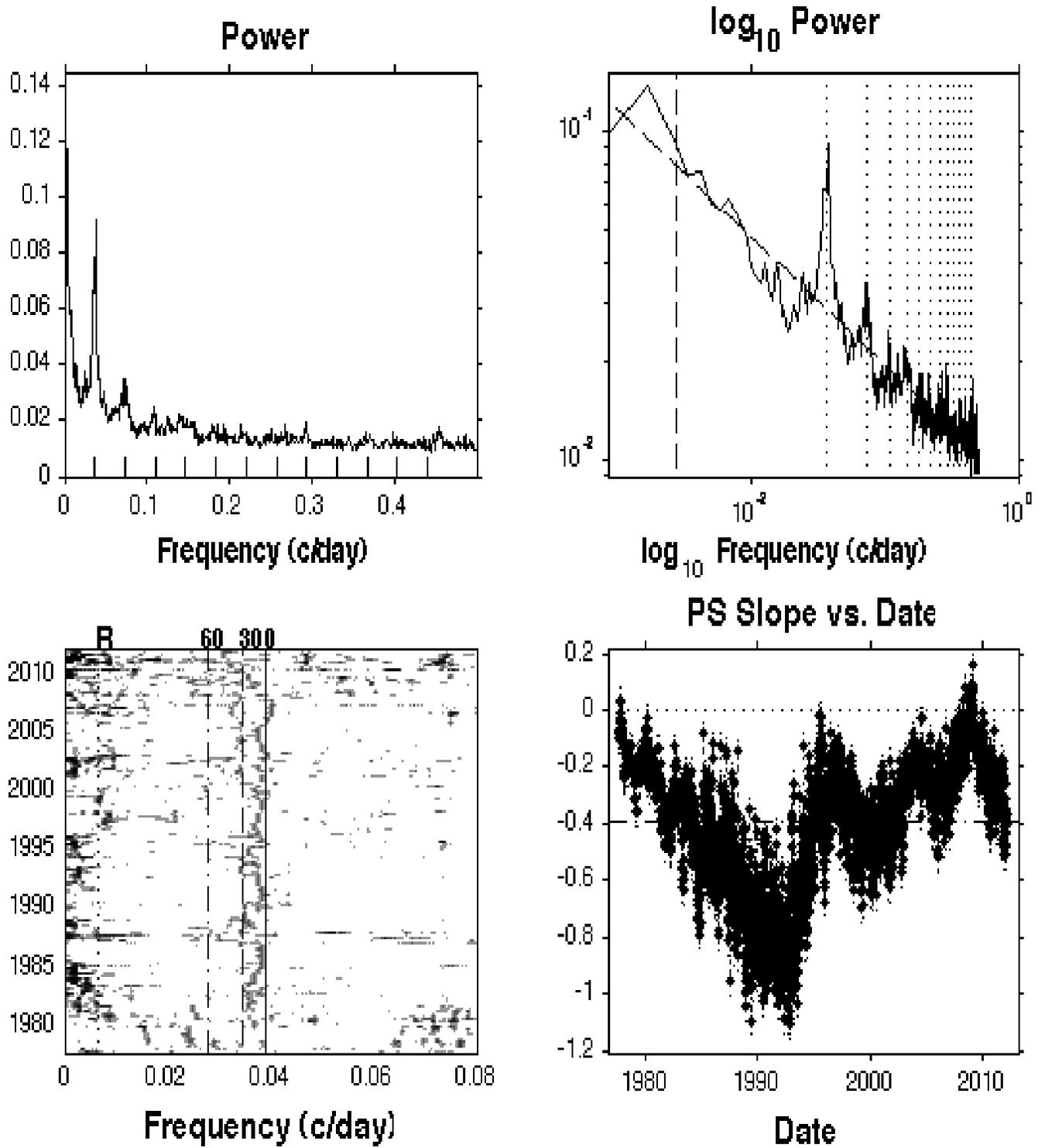} 
   \caption{\baselineskip 16pt Power spectra (with power-law fits) and time-frequency distributions of \emdx.}
\label{fig_mix_1}
\end{figure}
\clearpage

\begin{figure}[htb]
    \includegraphics[width=7in,height=8in]{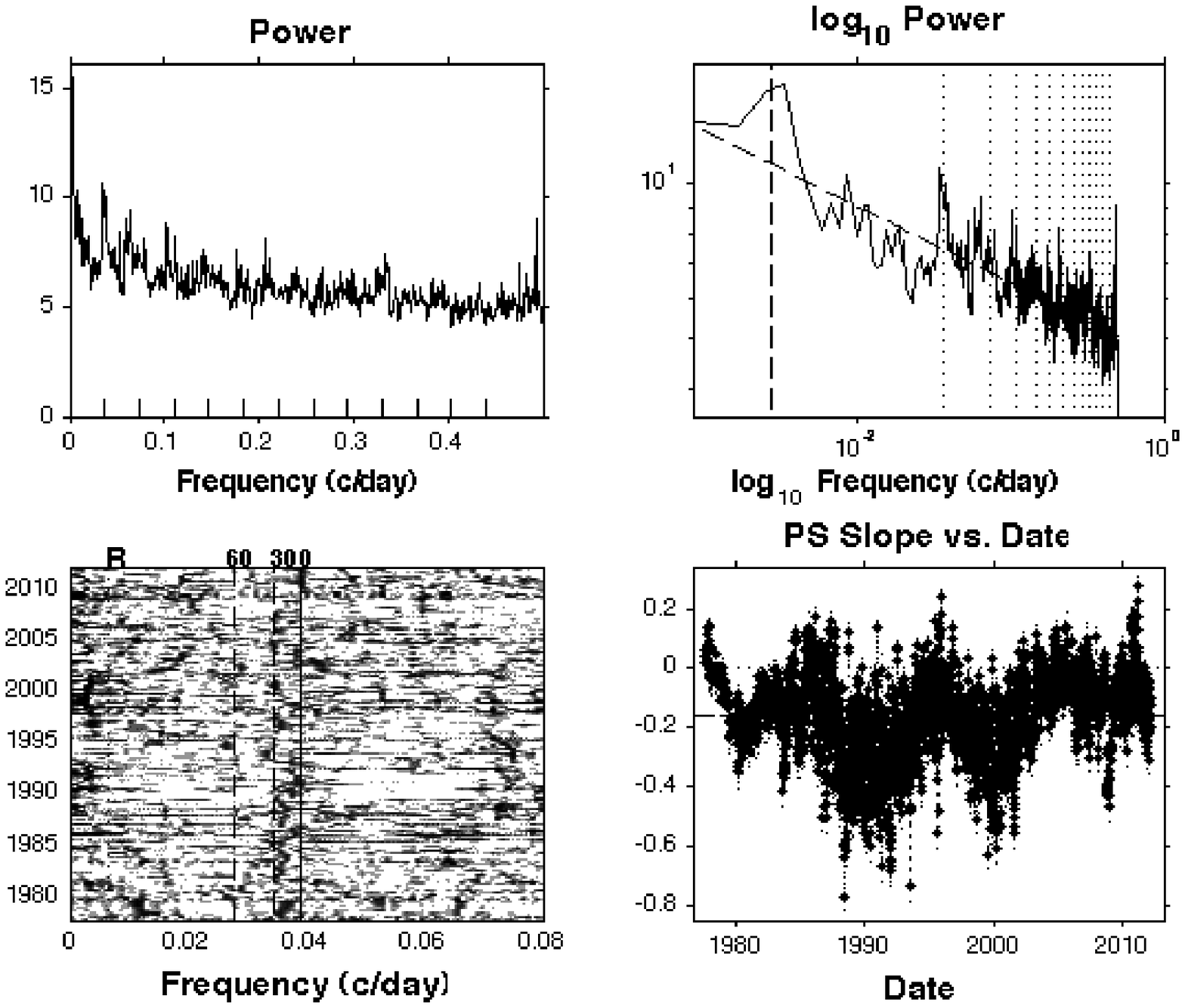} 
   \caption{\baselineskip 16pt Power spectra (with power-law fits) and time-frequency distributions of  \viored.}
\label{fig_mix_2}
\end{figure}
\clearpage

\begin{figure}[htb]
    \includegraphics[width=7in,height=8in]{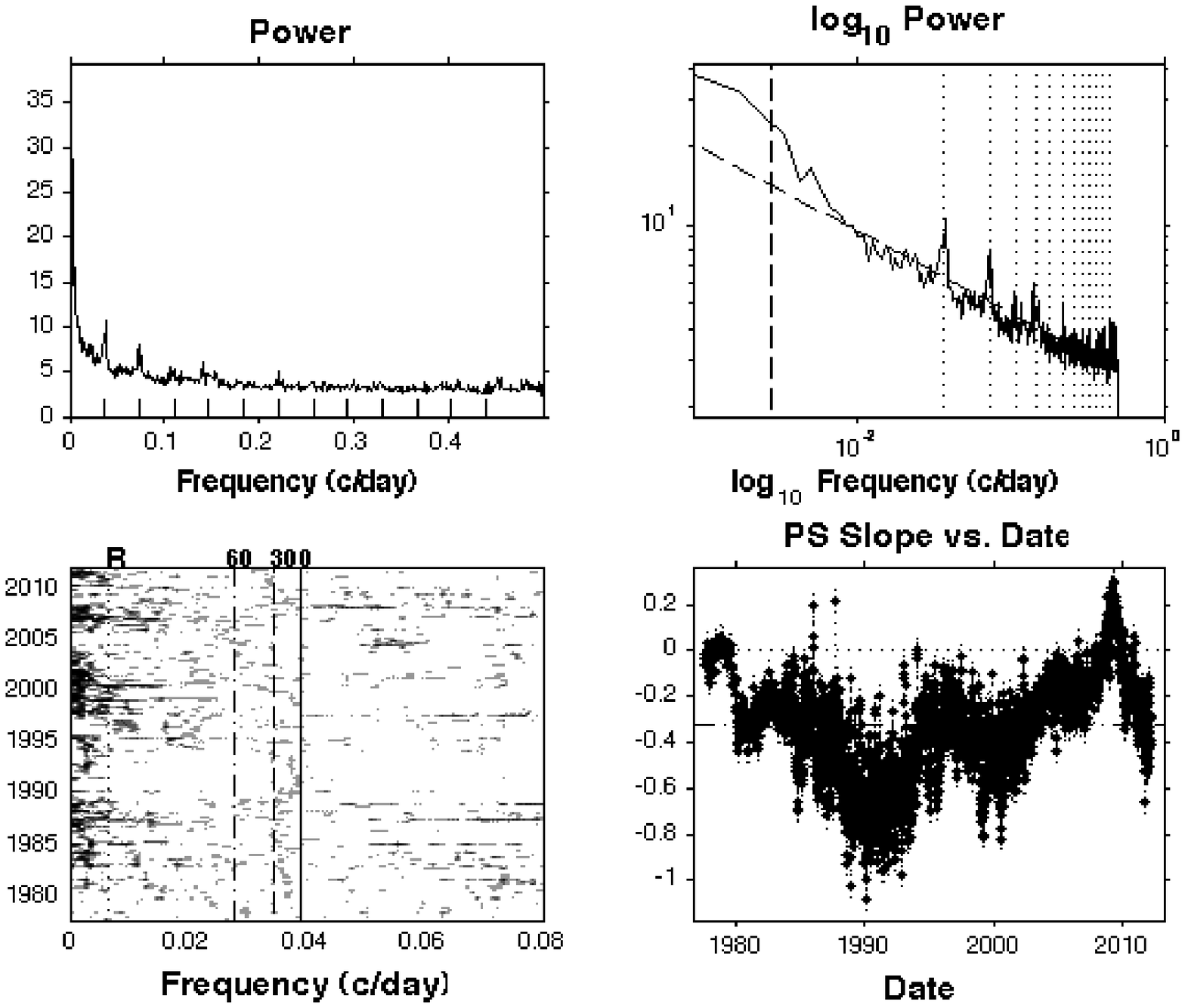} 
   \caption{\baselineskip 16pt Power spectra (with power-law fits) and time-frequency distributions of  \kvk.}
\label{fig_mix_3}
\end{figure}
\clearpage

\begin{figure}[htb]
    \includegraphics[width=7in,height=8in]{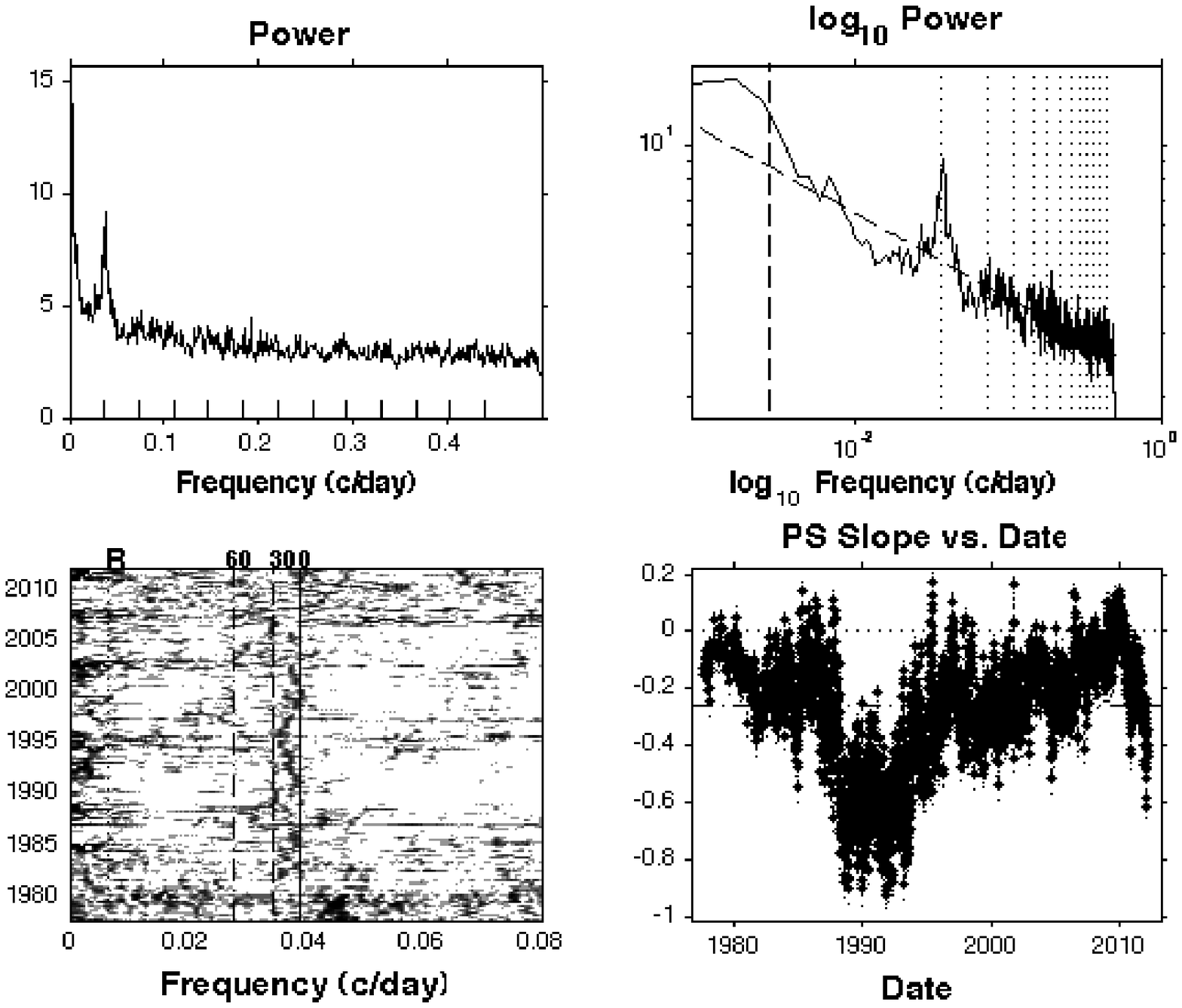} 
   \caption{\baselineskip 16pt Power spectra (with power-law fits) and time-frequency distributions of  \delkmax.}
\label{fig_mix_4}
\end{figure}
\clearpage

\begin{figure}[htb]
    \includegraphics[width=7in,height=8in]{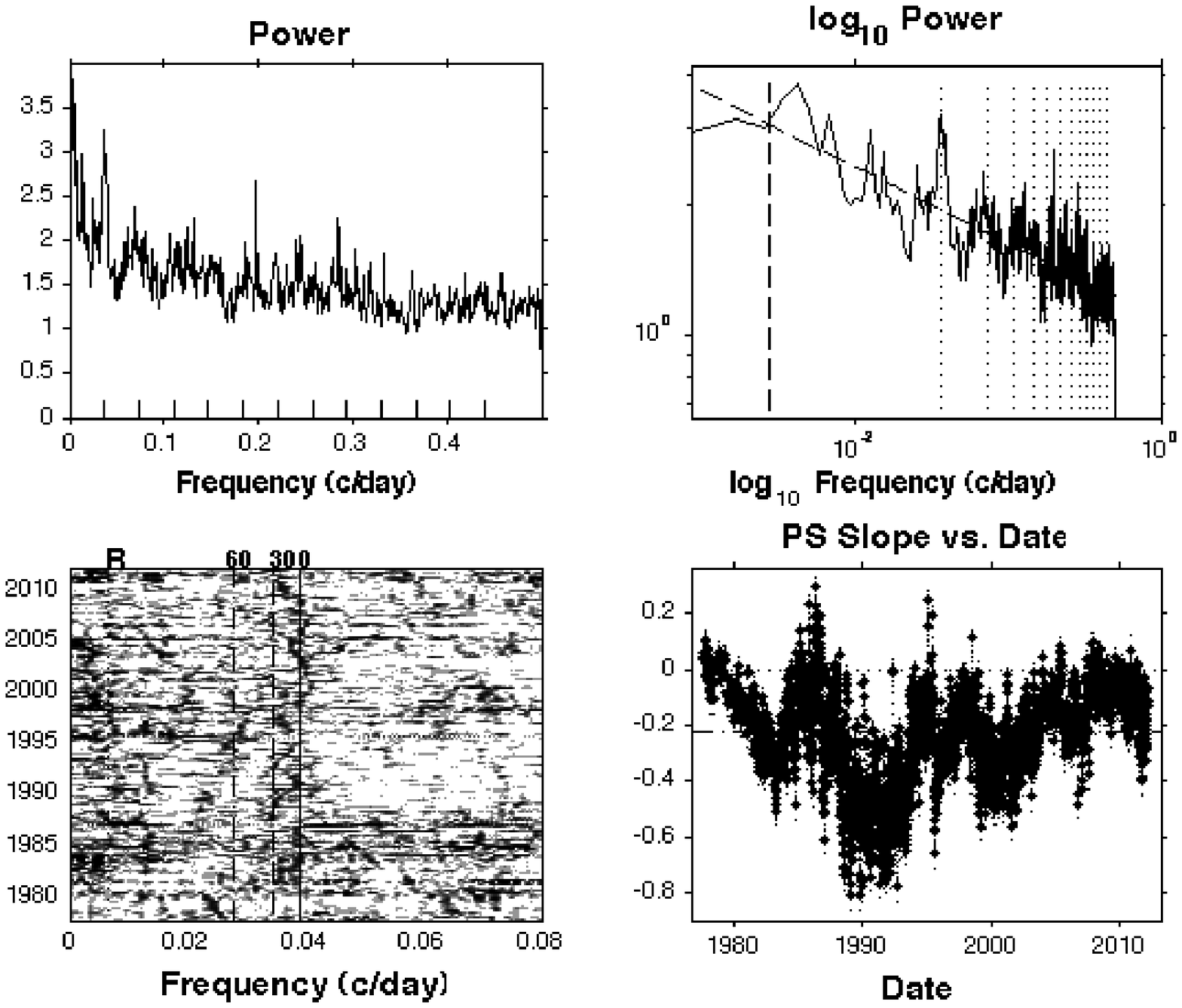} 
   \caption{\baselineskip 16pt Power spectra (with power-law fits) and time-frequency distributions of  \delkmin.}
\label{fig_mix_5}
\end{figure}
\clearpage

\begin{figure}[htb]
    \includegraphics[width=7in,height=8in]{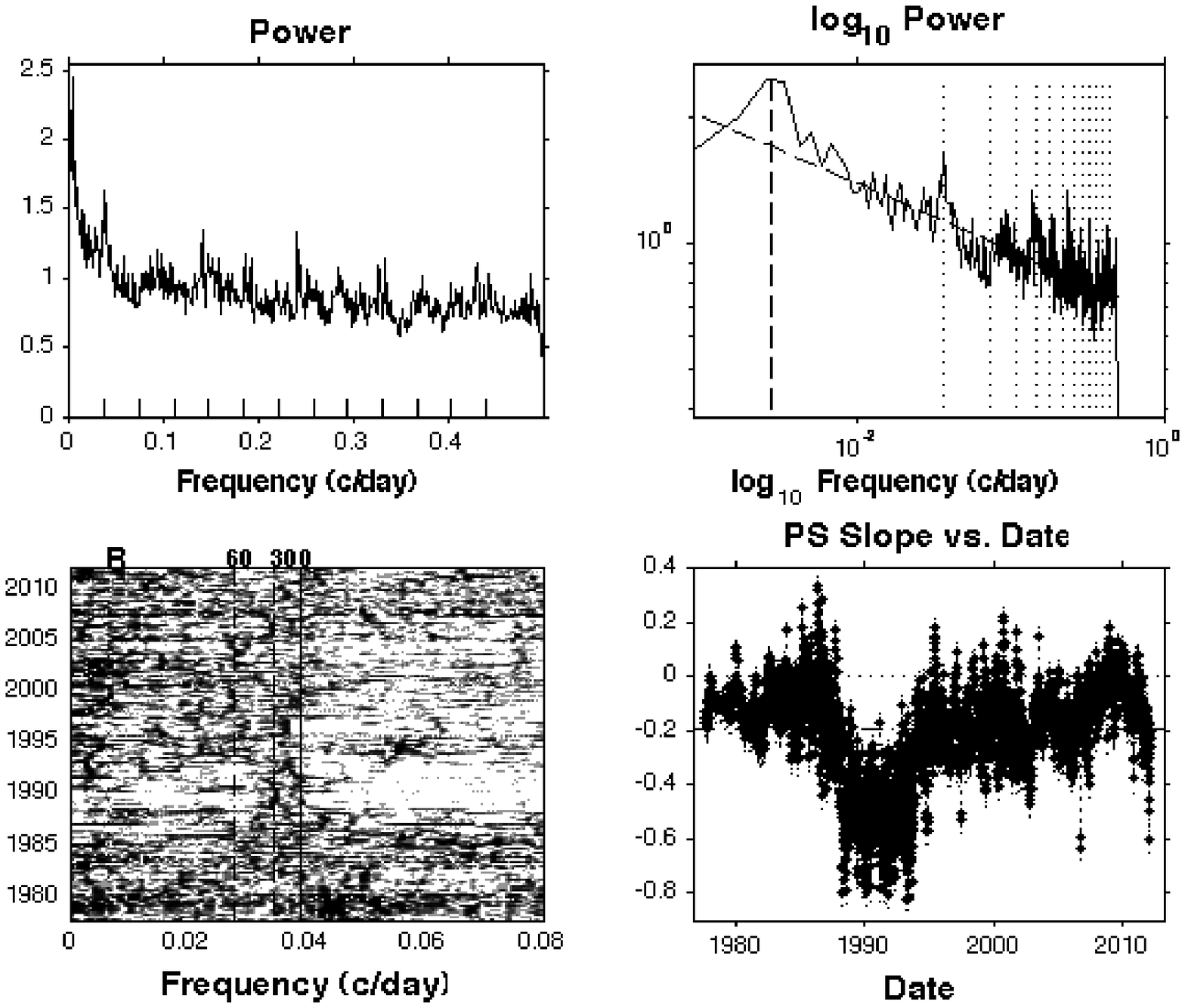} 
   \caption{\baselineskip 16pt Power spectra (with power-law fits) and time-frequency distributions of \delwb.}
\label{fig_mix_6}
\end{figure}
\clearpage

\begin{figure}[htb]
    \includegraphics[width=7in,height=8in]{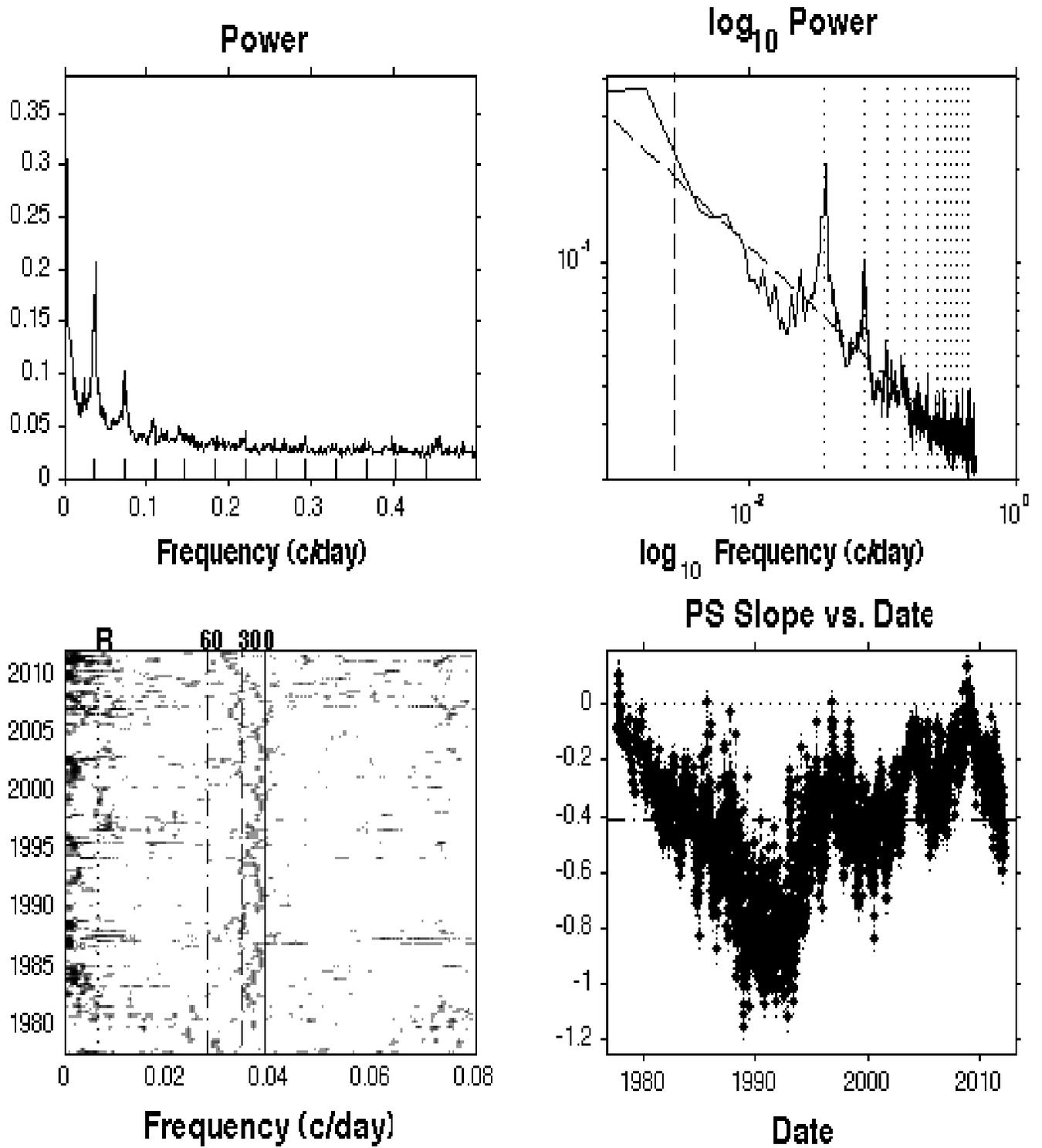} 
   \caption{\baselineskip 16pt Power spectra (with power-law fits) and time-frequency distributions of \kthree.}
\label{fig_mix_7}
\end{figure}
\clearpage

\begin{center}
Appendix 2: Notes on the Computations
\end{center}

\begin{figure}[htb] 
    \includegraphics[width=6.75in,height=4in]{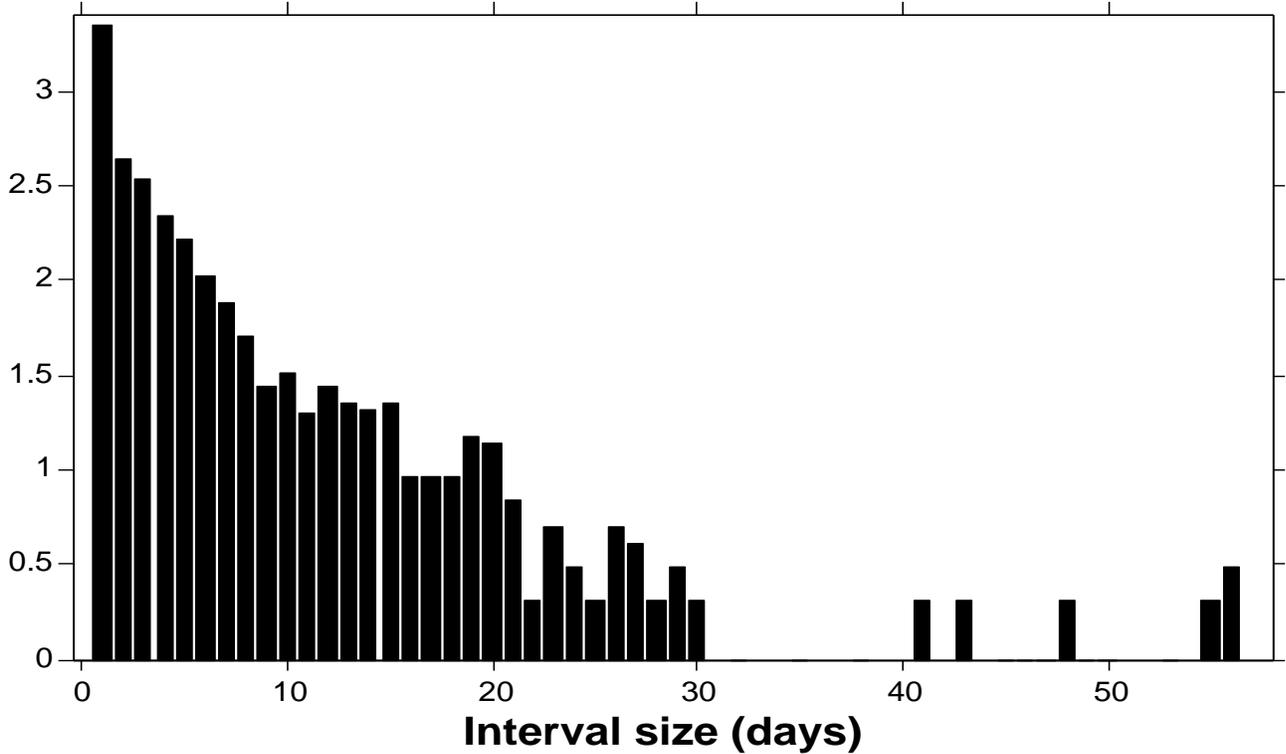} 
   \caption{\baselineskip 16pt Distribution of data gaps: Logarithm (base 10) 
   of the 
   number vs. interval size.
   For clarity this histogram
   omits nine large gaps 
   (62, 70,  71, 82, 83, 123, 130,  216 and 174 days)
   from the early years of the program (before 1982.5).
   The spike of one-day intervals (2125 out of 3732, or 57 per cent )  
   refers to the nominal daily sampling;
   anything larger is a gap. }
   \label{gaps}
\end{figure}

These time series represent 
a special case of irregular sampling, namely
evenly spaced (at 1-day intervals) but 
with some missing observations.
The degree of 
departure from 
even sampling is 
depicted in Figure \ref{gaps}.

Because of the non-trivial number 
of gaps and the wide distribution of their sizes 
it is necessary to compute correlation 
functions and power spectra with methods
that account for uneven sampling.
For direct computation of frequency domain
quantities 
(\emph{e.g.} Fourier transforms, phase and power spectra)
the Lomb-Scargle Periodogram 
\citep{vanicek,wright,lomb,scargle_ii,scargle_iii} 
is used here,
and has been previously used
to study this very data \citep{donahue_keil}.
This appendix describes the computation of
correlation functions for arbitrarily spaced data
for their own sakes as well as as an
alternative route to frequency domain quantities.

These computations start with the correlation
algorithm \citep{edelson} often used
in astronomy and well studied in the signal
processing literature, under the name of
\emph{slotted techniques} -- 
see e.g. \citep{stoica,rehfeld}.
It is an effective way to estimate the auto-correlation
function for unevenly spaced time series data 
$x(t_{n})$ 
such as we have here.  The basic idea is 
to construct
bins in the lag variable $\tau$, and then
sum the product $x(t_{1}) x(t_{2})$ 
over all data pairs such that the difference
 $t_{1} - t_{2}$ lies in a given such bin:
 \begin{equation}\label{tau_condition}
\tau_{k}  \le  t_{1} - t_{2} \le \tau_{k} + \Delta\tau \ ,
\end{equation}
\noindent
where $\tau_{k} $ denotes the start of bin $k$
and $\Delta\tau$ is the bin width.
That is to say the autocorrelation estimate is 
 \begin{equation}\label{eandk}
\rho( \tau_{k} ) 
= {1 \over N_{k} } \sum_{n} x_{n} x_{m} \ ,
\end{equation}
\noindent
where the sum is over all pairs $n,m$ such that
the corresponding time difference $t_{n} - t_{m}$
lies within the bin 
defined in Eq. (\ref{tau_condition}),
and $N_{k}$ is the number of terms in the sum.
It is more usual to write this formula replacing $x_{n}$ with
$x_{n} - \mu_{x}$,
where $\mu_{x}$ is the mean value of $x$,
either theoretical or empirical. Here we assume
an empirical mean has been subtracted.
The basic idea is that the average product
$x(t_{1}) x(t_{2})$ describes the degree to which
values separated by $\tau$ are related 
(large if positively correlated, 
large and negative if anti-correlated,
and small if uncorrelated).

The role of the factor ${1 / N_{k} }$
is interesting.  In estimating correlation
functions for evenly spaced data
two variants are used
\begin{equation}\label{even_acf_n}
\rho( k )  = {1 \over N } \sum_{n} X_{n} X_{n+k} \\
\end{equation}
\noindent
and
\begin{equation}\label{even_acf_nk}
\rho( k )  = {1 \over N  - k } \sum_{n} X_{n} X_{n+k} 
\end{equation}
\noindent
representing a trade-off favoring 
small variance (with larger bias)
or unbiased (but with larger variance) respectively.
Equation (\ref{eandk}) corresponds to equation
(\ref{even_acf_nk}) in that in both cases 
the denominator in the prefactor is the
number of terms contributing to that value
of the lag, so the expression is truly an
average.  If desired the analog of 
Equation (\ref{even_acf_n}) could
be implemented in an obvious way.

Even though Equation (\ref{eandk})
seems a bit abstract it is easily computed
in practice.  For evenly spaced data with 
gaps the binning in $\tau$ should correspond
to the constant sampling interval.
The power spectrum can then be computed using
the well-known identity that the power spectrum 
is the Fourier transform of the autocorrelation
function (which needs to be evaluated to the
maximum lag possible, namely equal to the
entire time-span of the observations).
With even spacing of the lag variable
this transformation can be rapidly 
carried out using the fast Fourier transform.
One potential difficulty is the possibility that 
the sampling and choice of binning in $\tau$ 
yields some empty bins -- no terms in
Equation (\ref{eandk}) that satisfy 
Equation (\ref{tau_condition}).
With the sampling in the present case 
(and with $\Delta\tau = 1$ day)  
there are no empty bins.
In addition the power computed in
this way can be negative,
since the autocorrelation may lack
the properties that guarantee 
a non-negative Fourier transform.
In practice this is a small effect
ameliorated simply by taking the absolute
value.

With an algorithm in hand to compute the
power spectrum (either the procedure just
outlined or the Lomb-Scargle periodogram) 
it is completely straightforward to 
compute the time-frequency distribution
simply by accumulating a matrix of power
spectra of the data points in a sequence of windows
slid along the observation interval.
The most important parameter is the width of
the window.  A good choice with the present
data was found to be about 0.05 times the
whole interval, or about 1.8 years.  

\noindent



\begin{thebibliography}{99}


\bibitem[\protect\citeauthoryear{Bai and Sturrock} {1993}]{bai}
Bai, T. and Sturrock, P. 1993, 
\apj, 409, 476.

\bibitem[\protect\citeauthoryear{Brevdo} {2009}]{brevdo}
Brevdo 2009,
MatLab toolbox for Synchrosqueezing:
\verb+http://www.math.princeton.edu/~ebrevdo/synsq/+.

\bibitem[\protect\citeauthoryear{Brown et al.} {1989}]{brown}
Brown, T. M., 
Christensen-Dalsgaard, J.,
Dziembowski, W. A.,
Goode, P., Gough, D. O.,
Morrow, C. A. 1989,
\apj, 343, 526-546.


\bibitem[\protect\citeauthoryear{Daubechies et al.} {2009}]{daubechies}
Daubechies, I.,  Lu, J.  and Wu, H.-T. 2011, 
Applied and Computational Harmonic Analysis,
30, 2011, 243

\bibitem[\protect\citeauthoryear{Domingo et al. } {2009}]{domingo}
Domingo, V., G
Ermolli, I.,
Fox, P.,
Haberreiter, M.,
Krivova, N.,
Kopp, G.,
Schmutz, W.,
Solanki, S. K.,
Spruit, H. C., Unruh,m Y.,
and V\"ogler 2009,
Space Sci. Rev., 145, 337

\bibitem[\protect\citeauthoryear{Donahue and Keil} {1995}]{donahue_keil}
Donohue, R., and Keil, S. 1995,
Solar Physics, 159, 53

\bibitem[\protect\citeauthoryear{Edelson and Krolik} { 1988}]{edelson}
Edelson, R. A. and Krolik, J. H. 1988,
Astrophysical Journal, 333, 646

\bibitem[\protect\citeauthoryear{Flandrin } {1999}]{flandrin}
Flandrin, P., 
Time-Frequency/Time-Scale Analysis 
(In French: Temps-FrŽquence) 1999,
Academic Press: London

\bibitem[\protect\citeauthoryear{Frasca et al. } {2011}]{frasca}
Frasca, A., Fršhlich, H.-E., Bonanno, A., Catanzaro,  G., 
Biazzo, K. and Molenda- ú Zakowicz, J. 2011,
Astronomy and Astrophysics, 532, A81

\bibitem[\protect\citeauthoryear{Georgieva } {2011}]{georgieva}
Georgieva, K., ``Why the Sunspot Cycle is Double Peaked,''
ISRN Astronomy and Astrophysics (2011), Article ID 437838
arXiv:1103.4552
	
\bibitem[\protect\citeauthoryear{Gonzalez et al.} {1990}]{gonzalez}
Gonzalez, W., Gonzalez, A., and Tsurutani, B. 1990,
 Planetary and Space Science, 
38, 181-187. 

\bibitem[\protect\citeauthoryear{Gottlieb et al.} {1975}]{wright}
Gottlieb, E. W., Wright, E. W. and Liller, W. 
1975, Astrophysical Journal Letters, 195, L33 

\bibitem[\protect\citeauthoryear{Hall } {2008}]{hall}
Hall, J. C. 2008,
Stellar Chromospheric Activity. Living Reviews in Solar Physics, 5, 2-+. Retrieved from \verb+http://adsabs.harvard.edu/cgi-bin/nph-bib_query?bibcode=2008LRSP....5....2H+


\bibitem[\protect\citeauthoryear{Hathaway } {2010}]{hathaway}
Hathaway, D. 2010,
 ``The Solar Cycle,''
  \emph{Living Rev. Solar Phys.}, 
{\bf 7}, 1. 
\verb+http://www.livingreviews.org/lrsp-2010-1+

\bibitem[\protect\citeauthoryear{Hill et al. } {2009}]{hill}
Hill, M., Hamilton, D., and Krimigis, S. 2009,
\emph{Journal of Geophysical Research},
{\bf 106}, 8315-8322.


\bibitem[\protect\citeauthoryear{Howe } {2009}]{howe}
Rachel Howe, 
``Solar Interior Rotation and its Variation,''
Living Rev. Solar Phys. 6,  2009,  1.
http://www.livingreviews.org/lrsp-2009-1


\bibitem[\protect\citeauthoryear{Jackson et al. } {2005}]{jackson}
Jackson, B., Scargle, J.D., Barnes, D.,
Arabhi, S., Alt, A., Gioumousis, P., 
Gwin, E., Sangtrakulcharoen, P., Tan, L.,
and Tun Tao Tsai, 2005,
IEEE Signal Processing Letters,
12, 105- 108



\bibitem[\protect\citeauthoryear{Joshi and Joshi } {2005}]{joshi}
Joshi, B., and Joshi, A. 2005,
Solar Physics, 226, 153-161.

\bibitem[\protect\citeauthoryear{Keil et al. } {2011}]{nso_www}
Ca II K-line Monitoring Program, 
Keil, S., Henry, T., White, D., and Livingston, B.,
\verb+http://nsosp.nso.edu/data/cak.pdf+.
This and other relevant documents, including
the data analyzed in this paper, are 
available at the National Solar Observatory
web site \verb+http://nsosp.nso.edu/data/cak_mon.html+,
the data being at \verb+ftp://ftp.nso.edu/idl/cak.parameters+.


\bibitem[\protect\citeauthoryear{Keil and Worden} { 1984}]
{keil_worden}
Keil, S., Worden, S. P. 1984,
\apj, 276, 766

\bibitem[\protect\citeauthoryear{Keil et al.} { 1998}]{keil_1}
Keil, S., Henry, T., and Fleck, B. 1998,
NSO/AFRL/Sac Peak K-line 
Monitoring Program,
Synoptic Solar Physics,
ASP Conference Series, 140, 
Balasubramaniam, Harvey and Rabin, eds.,
also at \verb+http://nsosp.nso.edu/data/cak_mon.html+.

\bibitem[\protect\citeauthoryear{Khramova et al.} {2002}]{khramova}
Khramova, M., Kononovich, E., and Krasotkin, S. 2002,
``Solar cyclicity: fine structure and forecasting,''
in \emph{Solar variability: from core to outer frontiers}. The 10th European Solar Physics Meeting, 9 - 14 September 2002, Prague, Czech Republic. Ed. A. Wilson. ESA SP-506, Vol. 1. Noordwijk: ESA Publications Division, ISBN 92-9092-816-6, 2002, p.145


\bibitem[\protect\citeauthoryear{Livingston et al.} {2007}]{livingston}
Livingston, W., Wallace, L., White, O. R., 
and Giampapa, M. S. 2007,
\apj, 657, 1137


\bibitem[\protect\citeauthoryear{Lomb} { 1976}]{lomb}
Lomb, N. R. 1976,
Astrophysics and Space Science, 39, 447

\bibitem[\protect\citeauthoryear{Lou et al. } {2003}]{lou}
Lou, Y., Wang, Y., Fan, Z., Wang, S., and Wang, J.
2003,
\mnras, 345, 809


\bibitem[\protect\citeauthoryear{Massi et al. } {2005}]{massi}
Massi, M., Niedh\"{o}fer, Carpentier, Y., and Ros, E. 2005,
A.\&A, 435, L1

\bibitem[\protect\citeauthoryear{Percival and Walden} { 1993}]{percival_walden}
Percival, D. B. and Walden A. T. 1993,
\emph{Spectral Analysis for Physical Applications: 
Multitaper and Conventional Univariate Techniques},
Cambridge University Press, Cambridge, UK


\bibitem[\protect\citeauthoryear{Rehfeld et al. } {2011}]{rehfeld}
Rehfeld, K., Marwan, N., Heitzig, J. and Kurths, J. 2011,
Nonlinear Processes in Geophysics, 18, 
389

\bibitem[\protect\citeauthoryear{Rieger et al.} { 1984}]{rieger}
Rieger, E., 
Share, G., 
Forrest, D., 
Kanbach, G.,
Reppin, C., and 
Chupp, E. (1984)
Nature,  312, 623

\bibitem[\protect\citeauthoryear{Scargle} { 1982}]{scargle_ii}
Scargle, J. D. 1982,
\apj, 263, 835

\bibitem[\protect\citeauthoryear{Scargle} { 1989}]{scargle_iii}
Scargle, J. D. 1989,
\apj, 343, 874


\bibitem[\protect\citeauthoryear{Scargle et al. } {2013}]{scargle_vi}
Scargle, J., Norris, J., Jackson, B.  and Chiang, J. 2013,
\apj, 764, 167

\bibitem[\protect\citeauthoryear{Schrijver and Zwaan } {2000}]{karel}
Schrijver, C. J. and Zwaan, C. 2000,
Solar and Stellar Magnetic Activity,
Cambridge University Press.

\bibitem[\protect\citeauthoryear{Silva-Valio } {2011}]{silva}
Silva-Valio, A., and Lanza, A. F. (2011), 
Astronomy and Astrophysics, 529,  A36.

\bibitem[\protect\citeauthoryear{Stoica and Sandgren } {2006}]{stoica}
Stoica, P. and Sandgren, N. 2006, 
Digital Signal Processing, 16,  712.

\bibitem[\protect\citeauthoryear{Sturrock } {1996}]{sturrock}
Sturrock, P. 1996,
"A Conjecture Concerning the Rieger and Quasi-Biennial
Solar Periodicities," 
\verb+astro-ph/9609150+

\bibitem[\protect\citeauthoryear{Sturrock et al. } {2013}]{sturrock_and}
Sturrock, P., Bertello, L., Fischbach, E., 
Javorsek II, D., Jenkins, J. H., 
Kosovichev, A.,  and Parkhomov, A. G.,
2013, 
Astroparticle Physics, 42, 62

\bibitem[\protect\citeauthoryear{Van'\v{c}ek} { 1971}]{vanicek}  
Van'\v{c}ek, P. 1971,
Astrophysics and Space Science, 12, 10

\bibitem[\protect\citeauthoryear{White et al.} { 1998}]{white}
White, O., Livingston, W., Keil, S., and Henry, T.,
Variability of the Solar Ca II K Line over the 22 Year Hale Cycle 
1998,
Synoptic Solar Physics,
ASP Conference Series, 140, 
Balasubramaniam, Harvey and Rabin, eds.,
also at \verb+http://nsosp.nso.edu/data/cak_mon.html+.

\end{thebibliography}
\end{document}